\documentclass[journal]{IEEEtran}

\usepackage[T1]{fontenc}
\usepackage[latin9]{inputenc}
\usepackage{textcomp}
\usepackage{subfig}
\usepackage{amsmath}
\usepackage{graphicx}
\usepackage[ruled,vlined]{algorithm2e}
\usepackage{bbm}
\usepackage{epstopdf}
\usepackage{amsmath}
\usepackage{amssymb}
\usepackage{cite}
\usepackage{balance}
\usepackage[colorlinks = true, linkcolor=blue, citecolor=blue,urlcolor=black]{hyperref}
\usepackage{soul}

\begin{document}
\title{Rate-distortion Balanced Data Compression for Wireless Sensor Networks}
\author{Mohammad~Abu~Alsheikh,~\IEEEmembership{Student~Member,~IEEE,}
Shaowei~Lin,\\
Dusit~Niyato,~\IEEEmembership{Senior~Member,~IEEE,}
and~Hwee-Pink~Tan,~\IEEEmembership{Senior~Member,~IEEE}
\thanks{A preliminary version of this paper has appeared in the proceedings of the 17th ACM International Conference on Modeling, Analysis and Simulation of Wireless and Mobile Systems, see~\cite{abu2014efficient}.}
\thanks{M.~Abu~Alsheikh is with the School of Computer Engineering, Nanyang Technological University, Singapore 639798 and also with the Sense and Sense-abilities Programme, Institute for Infocomm Research, Singapore 138632 (stumyhaa@i2r.a-star.edu.sg). S.~Lin is with the School of Engineering Systems and Design Pillar, Singapore University of Technology and Design, Singapore 487372 (shaowei\_lin@sutd.edu.sg). D.~Niyato is with the School of Computer Engineering, Nanyang Technological University, Singapore 639798 (dniyato@ntu.edu.sg). H.-P.~Tan is with the School of Information Systems, Singapore Management University, Singapore 188065 (hptan@smu.edu.sg).}}

\maketitle
\begin{abstract}
This paper presents a data compression algorithm with error bound guarantee for wireless sensor networks (WSNs) using compressing neural networks. The proposed algorithm minimizes data congestion and reduces energy consumption by exploring spatio-temporal correlations among data samples. The adaptive rate-distortion feature balances the compressed data size (data rate) with the required error bound guarantee (distortion level). This compression relieves the strain on energy and bandwidth resources while collecting WSN data within tolerable error margins, thereby increasing the scale of WSNs. The algorithm is evaluated using real-world datasets and compared with conventional methods for temporal and spatial data compression. The experimental validation reveals that the proposed algorithm outperforms several existing WSN data compression methods in terms of compression efficiency and signal reconstruction. Moreover, an energy analysis shows that compressing the data can reduce the energy expenditure, and hence expand the service lifespan by several folds.
\end{abstract}


\section{Introduction}

By 2020, 24~billion devices are expected to be interconnected over the Internet of things (IoT) technology in which wireless sensor networks (WSNs) form an intrinsic operational component~\cite{gubbi2013internet}. In these large-scale sensing networks, data compression is required for encoding the data collected from sensors into fewer bits, and hence reducing energy and bandwidth consumption. However, the computational burdens of the intended compression algorithms must be considered. Specifically, traditional data compression schemes from information and coding theory cannot be directly applied to a resource limited framework like WSNs as they are designed to optimize storage rather than energy consumption~\cite{razzaque2013compression}. Data compression enhances the functionality of WSNs in three main ways. Firstly, compression at cluster heads, gateways, or even within sensor nodes is one key ingredient in prolonging network lifetime~\cite{fasolo2007network,razzaque2013compression}. Secondly, archiving the sensing raw data over several years requires a tremendous amount of storage that ranges from terabytes to petabytes~\cite{gandhi2009gamps}. Thirdly, data compression increases the networking security by sending compressed data instead of the raw one. In particular, an intruder must fully access the data decompression procedure along with its parameters to reconstruct the raw data. The security and data privacy problem is receiving more attention especially in human-centric sensing and wireless body area networks~\cite{li2010data,al2012security}.

Once a deep understanding of monitored phenomena is achieved, the precise absolute readings of the sensors are not required, and extending the network lifespan is favored while collecting data within tolerable error margins~\cite{liang2011efficient}. Lossy data compression methods in WSNs are preferable over the lossless ones as they provide better compression ratio at lower computational cost~\cite{razzaque2013compression}. However, most traditional lossy data compression algorithms in WSNs lack an error bound guarantee mechanism due to the high computational demand of data decompression and reconstruction~\cite{razzaque2013compression}. Moreover, the complexity of the decompression routine becomes critical when the data destination is another resource-constrained node in the network. Thus, the computational complexity of data decompression is still an important concern.

The above discussion motivates the need for a solution that collectively supports the aforementioned design essentials. Briefly, our main contributions in this paper are as follows.
\begin{enumerate}
\item We propose a low-cost (both compression and decompression) lossy data compression technique with error bound guarantee. The routines for compression and decompression are implemented using only linear and sigmoidal operations. The compressed data features can be fed to machine learning algorithms~\cite{alsheikh2014machine} to automatically predict human activities and environmental conditions.
\item Unlike many conventional methods, our unified method is easily customized for both temporal and spatial compression. This allows the design of a uniform sensing framework that does not require many dedicated compression solutions, i.e., one for each application.
\item The proposed compression algorithm introduces a free level of security as an offline learned decompression dictionary is needed to recover the data. Other conventional data compression algorithms, such as~\cite{liang2011efficient,schoellhammer2004lightweight,quer2009interplay,sadler2006data}, lack this benefit as they are based on static procedures and do not use encoding dictionaries.
\end{enumerate}
Experiments on real world datasets show that the algorithm outperforms several well-known and traditional methods for data compression in WSNs. Furthermore, we show that the data compression using the proposed algorithm helps in reducing the data consumption in WSNs.

The rest of the paper is organized as follows. We first summarize related works in spatial and temporal compression of sensor data in Section~\ref{sec:related}. Section~\ref{sec:topologies} presents the problem formulation and describes some network topologies where  data compression is befitting. We then provide a mathematical overview on neural network autoencoders, and propose a compression algorithm that exploits data spatial and temporal correlations using autoencoders while providing an error bound guarantee in Sections~\ref{sec:preliminaries}~and~\ref{sec:framework}, respectively. Then, we evaluate and discuss the performance of our algorithm in experiments with actual sensor data in Section~\ref{sec:experimental}. Finally, Section~\ref{sec:conclusion} concludes the paper by outlining our key results and potential future work.

\section{Related Work}\label{sec:related}

We identify a wide variety of coding schemes in the literature (e.g.,~\cite{Srisooksai201237,zordan2012compress,razzaque2013compression}) and discuss some important solutions for signal compression in WSNs in the following.

\subsection{Limitations of Conventional WSN Compression Methods}
The lightweight temporal compression (LTC) algorithm~\cite{schoellhammer2004lightweight} is a simple method to compress the environmental data. LTC is a linear method that represents a time series readings by using a set of connecting lines. A similar model-based approach is the piecewise aggregate approximation (PAA) algorithm~\cite{keogh2001locally} that reduces the dimensionality of source data by generating series of discrete levels. On the negative side,  both LTC and PAA are less efficient when the data values change significantly over time even if the data periodically follows the same pattern and values. Moreover, they can only be used for temporal data compression as their use for spatial compression is usually inefficient. 

Principal component analysis (PCA), also known as the Karhunen-Loeve transform, has been widely used to extract (linear) correlations among sensor nodes (e.g.,~\cite{gastpar2006distributed,shen2009adaptive,rooshenas2010reducing,malik2011methodology}). Furthermore, a major scheme in the development of lossy data compression relies on the transformation of raw data into other data domains. Examples of these methods include discrete Fourier transform (DFT), fast Fourier transform (FFT)~\cite{zordan2012compress}, and the different types of discrete cosine transforms (DCT)~\cite{quer2009interplay}. Naturally, these transformation methods exploit the knowledge of the application to choose the appropriate data domain that discards the least data content. However, such algorithms suffer from low performance when used to compress data spatially or when noises are present in the collected readings.

\subsection{Limitations of Compressive Sensing (CS)}\label{sec:CS_limitations}
On the condition that a sparse representation\footnote{A signal representation is considered sparse if it contains most or all information of the original signal using relatively small number of nonzero components.} of a given signal is achievable, compressive sensing (CS) can efficiently transform the signal into a compressed form, which will be used later to recover an approximation of the original signal. In~\cite{bajwa2006compressive,luo2009compressive,xiang2011compressed,quer2012sensing}, the adoption of compressive sensing in WSNs is presented. Applying CS in WSNs has many limitations. Firstly, the assumption of input signal sparsity is strong and requires careful consideration in real-world deployments. Specifically, WSN data may not be sparse in the conventional data representations, such as the time, wavelet, and frequency domains. Similarly, linear sparse coding methods, such as dictionary learning~\cite{mairal2009online}, result in poor reconstruction fidelity due to the typical nonlinear correlations in WSN data. Secondly, introducing a few noisy readings may corrupt the sparse data representation (e.g., this issue is shown in~\cite{luo2009compressive} for the DCT method). Thirdly, CS requires the transmission of 3-4~times the number of nonzero values in the original signal for effective signal recovery~\cite{luo2009compressive}. This can result in low compression performance in  real-world WSN deployments. Finally, the complexity of CS data decompression hinders the development of error bound guarantee for CS-based compression methods in WSNs.


To address these limitations of existing methods, we propose a low-complexity lossy compression algorithm that exploits the nonlinear spatial-temporal correlations in WSN data and provides an error bound guarantee. Our algorithm automatically discovers intrinsic features in WSN data for efficient data representation, rather than relying on given features which may not suit the data well.

\section{System Model and Problem Formulation}\label{sec:topologies}

Throughout the paper, we will use the following standard notational conventions: A matrix is denoted by a bold uppercase letter, a vector is represented by a bold lowercase letter, and a scalar is denoted by a lowercase letter. Finally, functions and constants are named by uppercase letters.

\begin{figure}
\begin{centering}
\includegraphics[width=0.95\columnwidth,trim=1cm 1cm 1cm 1cm]{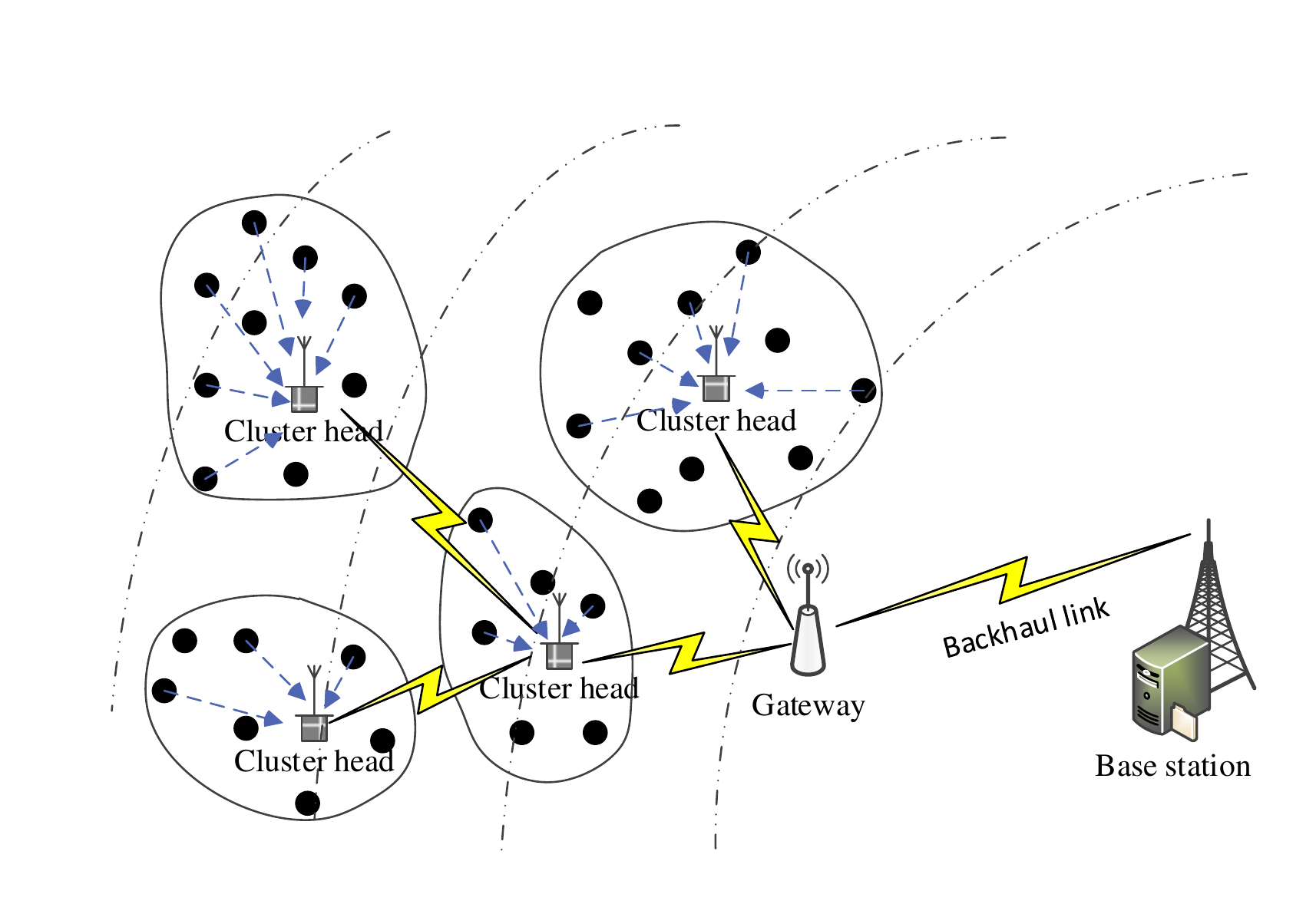}
\par\end{centering}

\caption{\label{fig:clustered_architecture}System model for data aggregation and funneling application over a backhaul link. Data can be compressed at the sensors, cluster heads, or gateway.}
\end{figure}

In this section, we give an overview of the problem considered in this paper including the data compression schemes (i.e., spatial and temporal compression). As shown in Figure~\ref{fig:clustered_architecture}, assume that each observed sample of sensor $i$ ($i=1,\ldots,N$) at time instant $t$ ($t=1,\ldots,M$) is formed as
\begin{equation}
x_{i}[t]=x_{i}^{*}[t]+w_{i}[t],
\end{equation}
where $i$ is the spatial location identifier and $t$ is the discrete time identifier. Consequently, all sensors are assumed to be synchronized according to the discrete-time model. $N$ is the number of spatial locations which is equal to the number of sensor nodes (i.e., each location is covered by a single sensor). $x_{i}^{*}[t]$ is the noiseless physical phenomenon value (e.g., a temperature value), and the noise values $\left\{w_{i}[t]\thicksim\mathbb{N}(0,\sigma_{w}^{2})\right\}_{i=1}^{N}$ are i.i.d random Gaussian variables with zero mean and variance $\sigma_{w}^{2}$ that depends on the hardware accuracy. Moreover, we assume that $\varphi_1<\left|x_{i}[t]\right|<\varphi_2$ which is defined as the dynamic range of the sensors with $\varphi_1$ and $\varphi_2$ as constants that are usually given in hardware data sheets. Thereby, any sample value that falls outside this sensing range is considered as an outlier reading (e.g., generated by a defective sensor). For example, the RM Young wind monitoring sensor (model 05103)~\cite{rm_young05103} measures the wind speed in the range of 0 to 100 m/s. Therefore, any reading beyond this range is considered as invalid data and should be eliminated.

Naturally, compression algorithms exploit the redundancy to extract spatial and temporal correlations from data. The choice of an optimal data compression scheme for a WSN is affected by network topology, routing algorithm, and data patterns~\cite{duarte2012signal}. To simplify the notations, we will consider the data vector denoted by $\textbf{x}\in\mathbb{R}^{L}$, $L\in\left\{N,M\right\}$, that is formed from a single location's measurements over time (a temporal data vector from a single sensor) or by combining many locations' measurements at a single time instant (a spatial data vector from many locations). Assuming perfect (without any missing or outlier values) data samples in $\textbf{x}$, the data compression (either temporal or spatial) is intended to represent $\textbf{x}$ in a compressed form $\textbf{y}\in\mathbb{R}^{K}$, where $K<L$. The compressed data is sent over a wireless channel to a base station (BS) over a backhaul link or using multihop transmissions. The BS must be able to compute a reconstruction of the original data $\hat{\textbf{x}}$. The reconstruction may be required to be within a guaranteed threshold (i.e., a tolerable error margin). Next, we give an overview of the data compression schemes considered in this paper and the network topologies that fit each scheme.

\subsection{Temporal (Intrasensor) Compression} 
This compression scheme exploits data redundancy of one sensor over time. Each sensor independently compresses its own data before transmission. Temporal compression is independent of the network topology as it does not require inter-sensor communication~\cite{duarte2012signal}. The temporal compression achieves maximum performance when the observed phenomenon changes slightly over time, such as hourly temperature or humidity readings. At a specific location~$i$, the temporal data vector is formed as $\textbf{x}=\left\{x_i[t]\right\}_{t=1}^{M}\in\mathbb{R}^{M}$. $M$ is designed to fit the physical phenomenon cycle and the sampling rate.

\subsection{Spatial (Inter-sensor) Compression}
In a dense network, data collected by neighboring sensors is highly correlated~\cite{pradhan2002distributed}. Spatial compression investigates the disseminated data patterns among different sensors over the area. Therefore, the performance of the spatial compression algorithm will be affected by the network topology and sensor deployment. Unlike temporal compression that considers only a single sensor, spatial compression considers a WSN with $N$ nodes. At a specific time instant~$t$, the samples of all sensors are used to construct a single data vector as $\textbf{x}=\left\{x_i[t]\right\}_{i=1}^{N}\in\mathbb{R}^{N}$.

\begin{figure}
\begin{centering}
\includegraphics[width=0.7\columnwidth,trim=1cm 1cm 1cm 0cm]{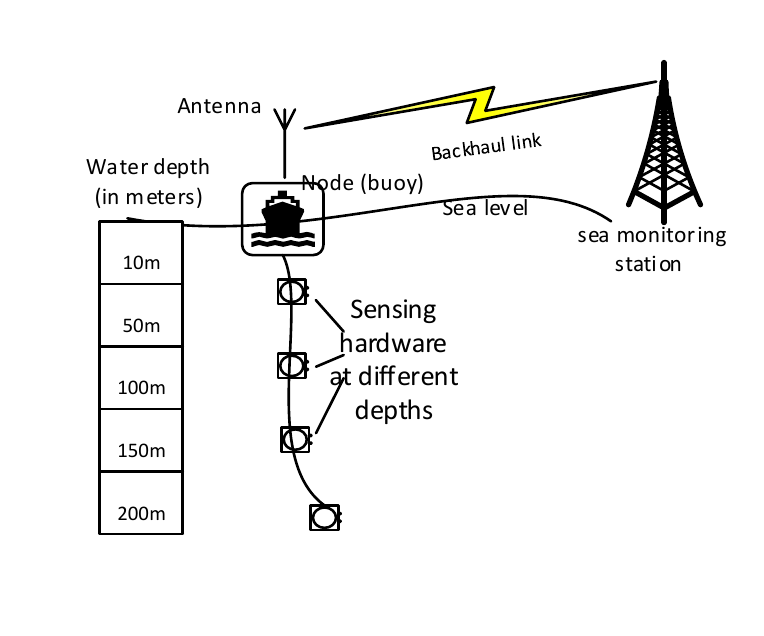}
\par\end{centering}
\caption{\label{fig:wireless_bouy}A wireless sensor node to monitor the water condition (e.g., pH level) at different depths using sensing components fixed by a cable to a buoy.}
\end{figure}

Another fitting architecture for spatial compression is demonstrated in Figure \ref{fig:wireless_bouy}. A buoy sensor node is used for data monitoring at different depths of the sea, where each node comprises a collection of sensors and a single transmission unit. The data is compressed at the buoy node by exploiting the spatial correlation among sensor's readings before transmitting them to a sea monitoring station.


The next section gives an overview of a special type of artificial neural networks called the autoencoder network. The discussion describes the procedure that is followed to generate a compressed data representation at a hidden layer and reconstructed data values at an output layer. Our algorithm will be later developed based on this formulation.

\section{Neural Autoencoders (AEs)}\label{sec:preliminaries}

Artificial neural networks (ANNs) have been successfully used in the development of novel solutions for WSNs as they can capture nonlinear structures in data~\cite{alsheikh2014machine}. For example, an ANN-based method for minimizing environmental influences on sensor responses was proposed in~\cite{patra2011development}. It has been shown in~\cite{abu2014area} that ANNs are a solid tool for maximizing sensing coverage. This paper presents an appealing application of ANNs for data compression in WSNs. The key technical challenges of this application are (i)~learning nonlinear spatio-temporal correlations of WSN data, (ii)~enabling low-cost data compression and decompression, (iii)~ensuring data reconstruction within tolerable error margins, and (iv)~minimizing WSN energy consumption.

\begin{figure}
\begin{centering}
\includegraphics[width=0.6\columnwidth,trim=1.2cm 0.8cm 1.2cm 0.5cm]{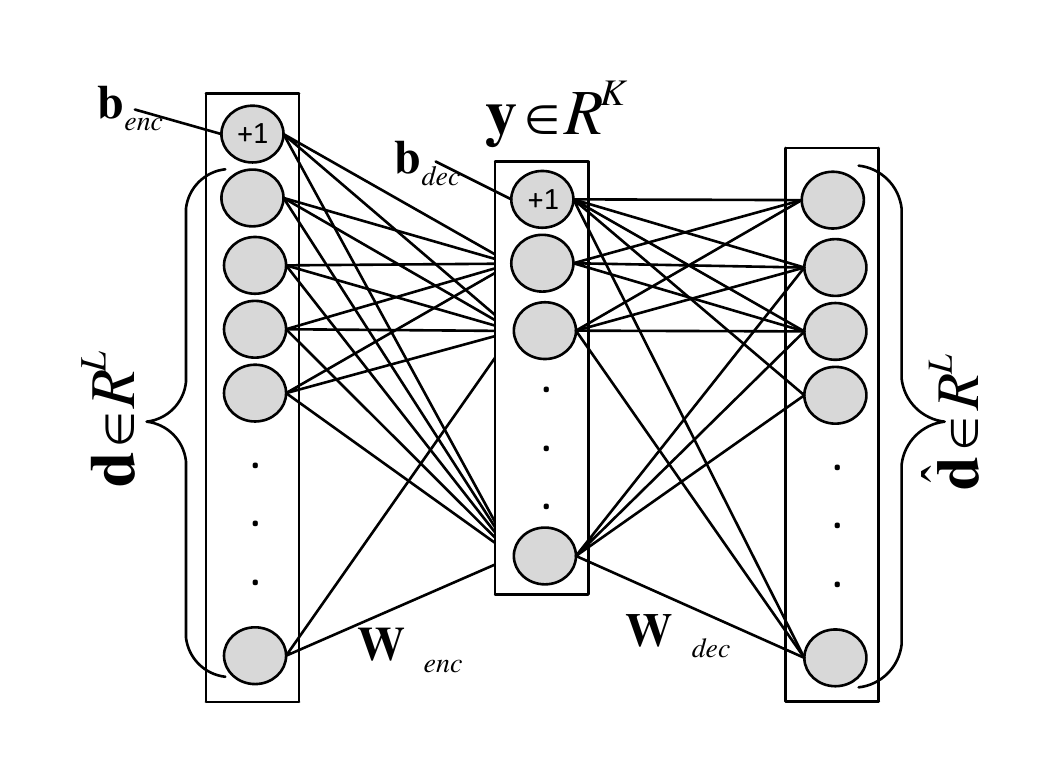}
\par\end{centering}
\caption{\label{fig:basic_autoencoder}Using AE to project the data to a lower dimensional representation ($K\ll N$).}
\end{figure}

An autoencoder (or auto-associative neural network encoder) is a three-layer neural network that maps an input vector $\mathbf{d}\in\mathbb{R}^{L}$ to a hidden representation $\mathbf{y}\in\mathbb{R}^{K}$ and finally to an output vector $\mathbf{\hat{d}} \in\mathbb{R}^{L}$ that approximates the input $\mathbf{d}$, as shown in Figure~\ref{fig:basic_autoencoder}. The vectors satisfy
\begin{subequations}
\begin{equation}
\mathbf{y}=F\left(\mathbf{W}_{enc}\mathbf{d}+\mathbf{b}_{enc}\right)
\end{equation}
\begin{equation}
\mathbf{\hat{d}}_{\boldsymbol{\theta} }(\mathbf{d})=F\left(\mathbf{W}_{dec}\mathbf{y}+\mathbf{b}_{dec}\right)
\end{equation}
\begin{equation}
F\left(\upsilon\right)=\frac{1}{1+\exp(-\upsilon)}
\end{equation}
\end{subequations}
where $\boldsymbol{\theta} :=[\mathbf{W}_{enc},\mathbf{b}_{enc},\mathbf{W}_{dec},\mathbf{b}_{dec}]$ are real-valued parameters that must be learned by a suitable training algorithm, and $F\left(\cdot\right)$ is the sigmoid function. Note that other nonlinear function such as the hyperbolic tangent can also be used. The parameters $\mathbf{W}_{enc}$ and $\mathbf{b}_{enc}$ are the encoding weight matrix and bias, while $\mathbf{W}_{dec}$ and $\mathbf{b}_{dec}$ are the decoding weight matrix and bias, respectively. The entries of $\mathbf{y}$ and $\mathbf{\hat{d}}$ are sometimes called activations.

To learn optimal neural weights $\boldsymbol{\theta}$ using training data $\mathbf{D}$, we define the cost function of the basic autoencoder (AE) as
\begin{equation}\label{eq:cost_function}
\begin{aligned}\Gamma_{\text{AE}}\left(\boldsymbol{\theta},\mathbf{D}\right) =\frac{1}{\vert\mathbf{D}\vert }\sum_{\mathbf{d}\in\mathbf{D}}{\frac{1}{2}\left\Vert \mathbf{d}-\mathbf{\hat{d}}_{\boldsymbol{\theta}}(\mathbf{d})\right\Vert ^{2}}.
\end{aligned}
\end{equation}
This function penalizes the difference between each input vector $\mathbf{d}$ and its reconstruction $\mathbf{\hat{d}}_{\boldsymbol{\theta} }(\mathbf{d})$. Consequently, the optimal neural weights may be computed using standard optimization algorithms such as the L-BFGS.

Different variants of the basic AE have been introduced in the literature to discourage the neural network from overfitting the training data~\cite{bengio2012practical}. Generally speaking, these regularization methods penalize the neural weight characteristics~or the hidden layer sparsity characteristics.

\textbf{Weight decaying autoencoder (WAE)}: In this variant, the cost function is defined with an extra weight decay term:
\begin{equation}
\Gamma_{\text{WAE}}\left(\boldsymbol{\theta},\mathbf{D}\right)= \Gamma_{\text{AE}}\left(\boldsymbol{\theta},\mathbf{D}\right)+
\frac{\alpha}{2}\left( \left\Vert \mathbf{W}_{enc} \right\Vert^2 +  \left\Vert \mathbf{W}_{dec} \right\Vert^2 \right)
\end{equation}
where $\left\Vert \mathbf{W} \right\Vert^2$ represents the sum of the squares of the entries of a matrix $\mathbf{W}$, and $\alpha$ is a hyperparameter%
\footnote{A hyperparameter is a variable that is selected a priori. This differentiates a hyperparameter from a model parameter (e.g., the encoding weight) which is adjusted during the learning process.%
} 
that controls the contribution from the weight decay term.

\textbf{Sparse autoencoder (SAE)}: This version extracts a sparse data representation at the hidden layer. In particular, we want most of the entries of $\textbf{y}$ to be close to zero. Sparsity is encouraged by adding the Kullback\textendash{}Leibler (KL) divergence function~\cite{ng2011sparse}:
\begin{subequations}
\begin{equation}
\Gamma_{\text{SAE}}\left(\boldsymbol{\theta},\mathbf{D}\right)=\Gamma_{\text{WAE}}\left(\boldsymbol{\theta},\mathbf{D}\right)+\beta\sum_{k=1}^{K}\text{KL}(\rho||\hat{\rho_{k}})
\end{equation}
\begin{equation}
\text{KL}(\rho||\hat{\rho_{k}})=\rho\log_{e}\frac{\rho}{\hat{\rho_{k}}}+\left(1-\rho\right)\log_{e}\left(\frac{1-\rho}{1-\hat{\rho_{k}}}\right)
\end{equation}
\end{subequations}
where $\beta$ is a hyperparameter that controls the sparsity weight, $\rho$ is the sparsity parameter (target activation) that is chosen to be close to zero, and $\hat{\rho_{k}}$ is the average activation of the $k$-th node in the hidden layer.

Next, the proposed algorithm is described in more details, and a discussion is provided to signify the advantages of our AE-based compression algorithm. Moreover, a method is presented to ensure data compression within a tolerable error margin (i.e., an error bound guarantee). Finally, simple but important methods for data preparation and missing data imputation are also presented.

\section{Lossy Compression with Error Bound Guarantee}\label{sec:framework}

\begin{figure}
\begin{centering}
\includegraphics[width=0.95\columnwidth,trim=1.2cm 1cm 1.2cm 0.0cm]{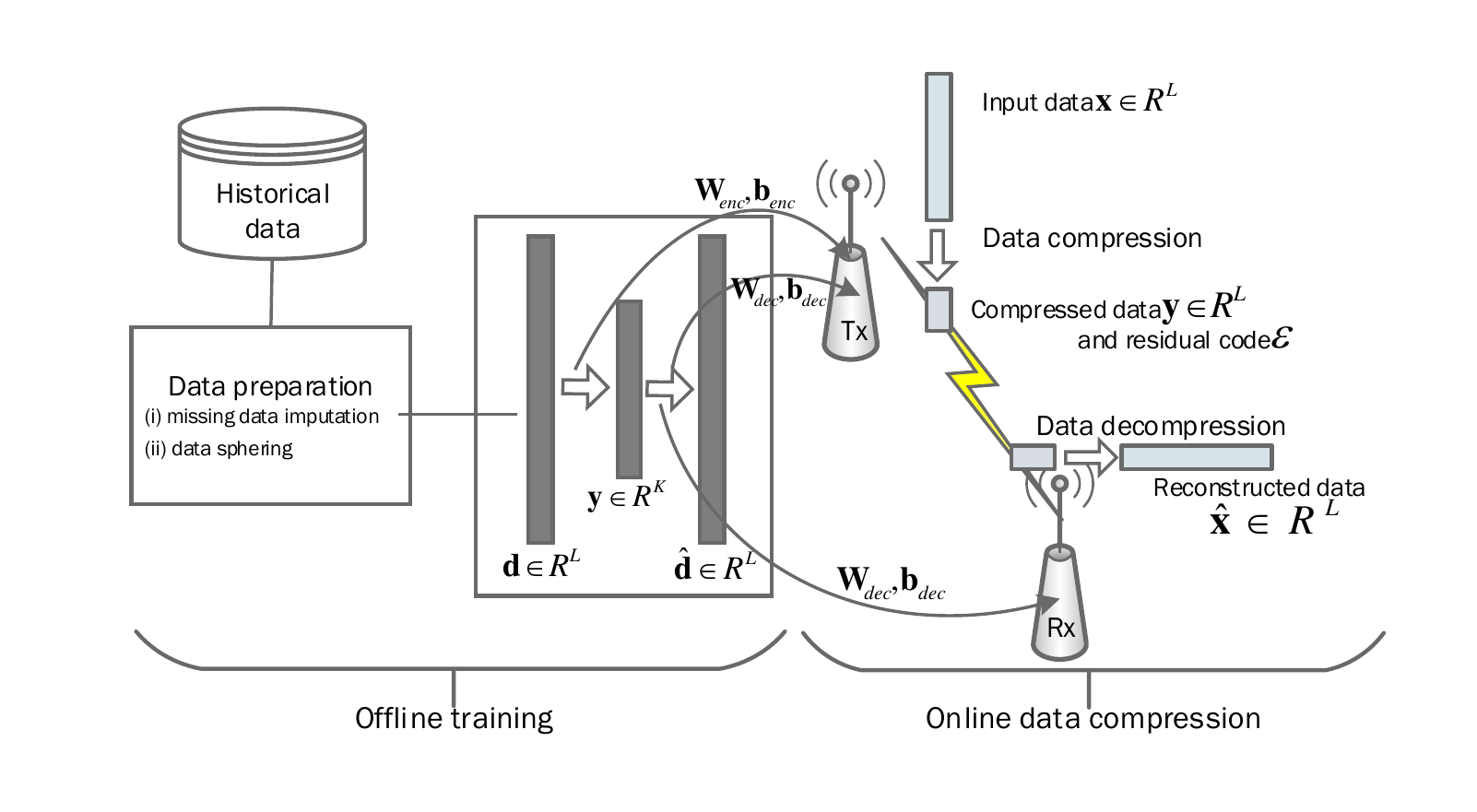}
\par\end{centering}

\caption{\label{fig:autoencoder_adoption}AE adoption for data compression and dimensionality reduction in WSNs. Initially, the parameters $\mathbf{W}_{enc},\mathbf{b}_{enc},\mathbf{W}_{dec},$ and $\mathbf{b}_{dec}$ are adjusted during the learning stage (offline mode). Subsequently, the encoding part will be executed in the transmitter side (Tx) to achieve a compressed representation of the data. Then the receiver (Rx) will deploy the decoding part to recover a proper approximation of the original signal.}
\end{figure}

We propose to apply the autoencoder to the data compression and dimensionality reduction problem in WSNs to represent the captured data using fewer bits as demonstrated in Figure \ref{fig:autoencoder_adoption}. The algorithm enables compressed data collection with tolerable error margins, and it contains three main steps: historical data collection using the sensor nodes, offline training and modeling at the BS, and online data temporal or spatial compression. The proposed algorithm is motivated by several reasons related to WSN characteristics, as well as the ability of AEs to automatically extract features in the data.
\begin{enumerate}
\item AEs are used to extract a suitable, low-dimensional code representation that retains most of the information content of the original data. Besides data compression, these intrinsic features are integral for data analytics and visualization algorithms~\cite{hinton2006reducing}, e.g., classification problems.
\item Sensor networks are deployed in a variety of distinct scenarios with different network structures and data patterns. The proposed algorithm has the flexibility of supporting many scenarios using one unified technique. 
\item Finally, after learning the AE's parameters, the process of data encoding and decoding are simple and can be programmed with a few lines of code.
\end{enumerate}

\subsection{Missing Data Imputation}\label{sec:missing_data}
\begin{figure}
\begin{centering}
\includegraphics[width=0.6\columnwidth,trim=1cm 0.8cm 1cm 0.8cm]{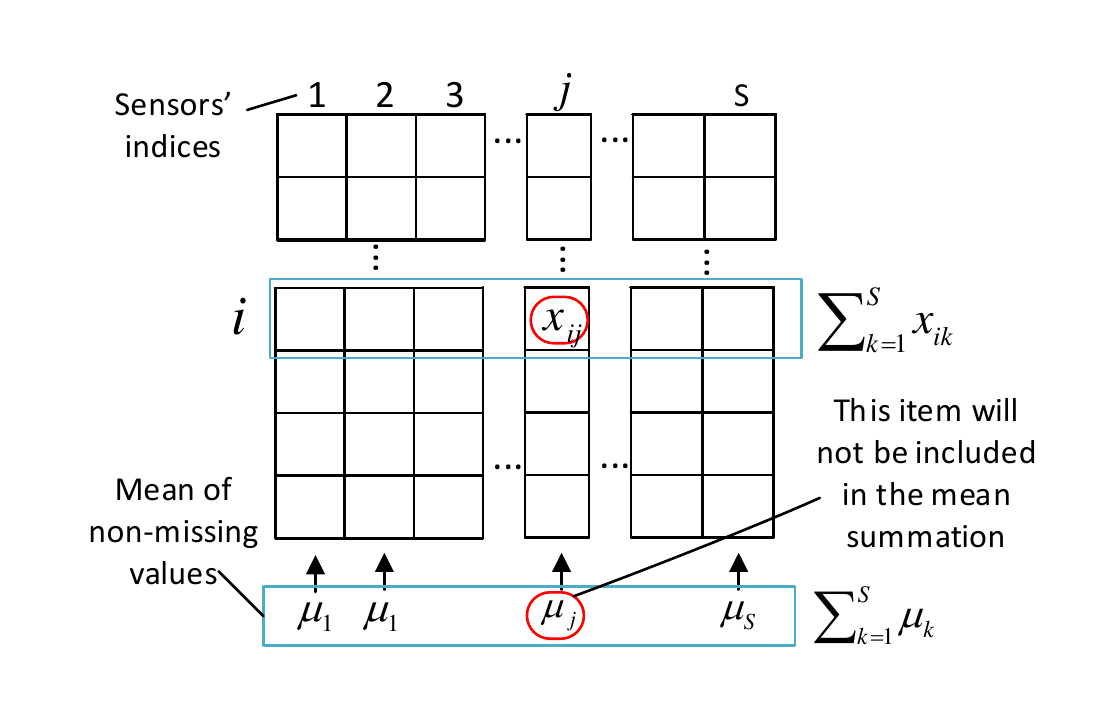}
\par\end{centering}
\caption{\label{fig:missing_data_prediction}Missing data prediction method.}
\end{figure}

Missing WSN data can occur due to wide variety of causes such as malfunctioning node, communication failure and interference, and unsynchronized sampling of the sensors. For missing data imputation, we use a simple naive method as shown in Figure \ref{fig:missing_data_prediction}. Suppose that the entry $x_{ij}$ in the aligned matrix is missing, where $i$ and $j$ are the time and sensor indices. Let $S$ be the set of observed sensors at time $i$, and let the mean of the observed readings of sensor $j$ be $\mu_{j}$. We estimate $x_{ij}$ as
\begin{equation}
\hat{x}_{ij}=\frac{\sum_{k \in S} x_{ik}}{\sum_{k\in S}\mu_{k}}\mu_{j}.
\end{equation}
In many sensor applications such as temperature monitoring, the naive method seems to work relatively well because of pseudo-linear correlations in the data. We chose this naive method because of the low computational resources available at the sensor nodes. 

\subsection{Data Sphering} 
The entries of the output vector $\mathbf{\hat{d}}$ of the AE are from the sigmoid function, so they are all between $0$ and $1$. Because the AE attempts to reconstruct the input vector $\mathbf{x} \in \mathbb{R}^L$, we need to normalize our input data so that the entries are also between $0$ and $1$. Moreover, for the AE to work, the input data vectors must be distributed somewhat uniformly near the unit sphere in $\mathbb{R}^{L}$. This process is called data sphering~\cite{ng2011sparse}. One simple method involves truncating readings that lie outside three standard deviations from the vector mean, and rescaling the remaining readings so that they are between $0.1$ and $0.9$. In particular, the formula is
\begin{equation} \label{eq:truncation}
\begin{split}
\mathbf{d} & = \text{normalize}(\mathbf{x},\sigma) \\
	&= 0.5 + \frac{0.4}{3\sigma} \max\left(\min\left(\mathbf{x}-\text{mean}(\mathbf{x}),3\sigma\right),-3\sigma\right)
\end{split}
\end{equation}
where $\mathbf{x}$ is the source data vector and $\sigma$ is the standard deviation of the entries of $\mathbf{x}-\text{mean}(\mathbf{x})$ over all $\mathbf{x}$ in the training dataset. $\textbf{d}$ is the data vector that is fed to the AE network. Furthermore, assuming the data is normally distributed, the probability that a reading is located within three standard deviations from the mean is $99.7\%$~\cite{kriegel2009loop}. Conversely, given the mean $m$, the original data vector $\mathbf{x}$ may be reconstructed (up to truncated outliers) as
\begin{equation} \label{eq:detruncation}
\begin{split}
\mathbf{p} & =\text{denormalize}(\mathbf{\hat{d}},m,\sigma) = \frac{3\sigma}{0.4} (\mathbf{\hat{d}}-0.5)+m,
\end{split}
\end{equation}
where $\mathbf{p}$ is a reconstruction of the source data $\textbf{x}$ by using the AE output vector $\mathbf{\hat{d}}$.

\subsection{Error Bound Mechanism}\label{sec:error_bound_mechanism}
The error bound $\epsilon$ is defined to be the maximum acceptable difference between each collected reading by the sensor and the recovered one by the receiver after receiving the compressed representation. Basically, the error bound is tuned by considering several factors such as the application requirements and the used sensors' precision. For example, the RM Young wind monitoring sensor (model 05103)~\cite{rm_young05103} measures the wind speed and direction with accuracy of 0.3 m/s and 3\textdegree{}, respectively. Thus, setting the error bound to be equal to the sensor accuracy may be an acceptable design basis.

\begin{figure}
\begin{centering}
\includegraphics[width=0.65\columnwidth,trim=1cm 1.0cm 1cm 0.5cm]{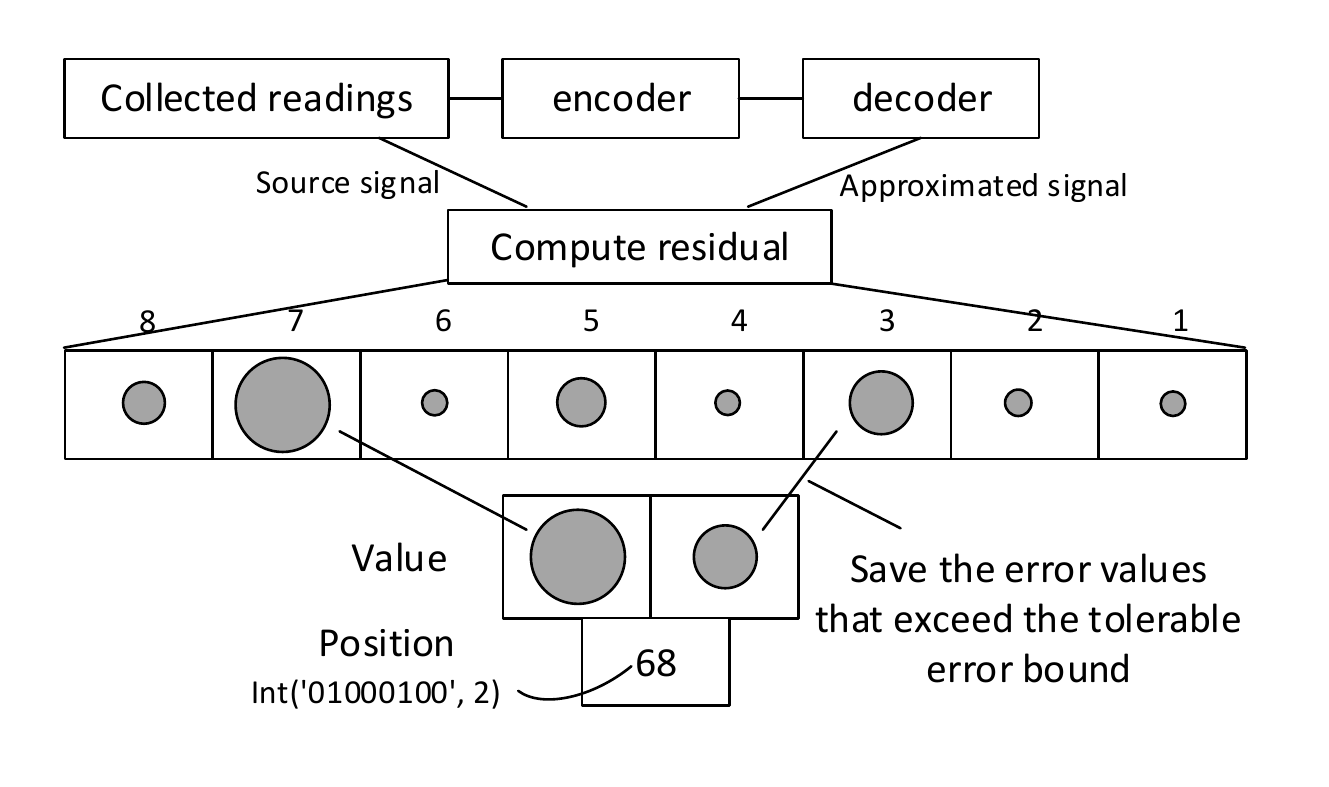}
\par\end{centering}

\caption{\label{fig:error_bound}The error bound mechanism performed by the transmitting node.}
\end{figure}

Let $\mathbf{p}$ be a reconstruction of $\textbf{x}$ that is not guaranteed to be within any tolerable error margin. The error bound mechanism first computes the residual $\mathbf{r} = \mathbf{x}-\mathbf{p}$ as shown in Figure \ref{fig:error_bound}. Any entry of the residual vector exceeding the bound $\epsilon$ will be transmitted, using the residual code
\begin{equation}
\boldsymbol{\varepsilon}=\text{residualCode}(\mathbf{r}, \epsilon)=\left(\mathbbm{1}_{J}, \left(r_{j}\right)_{j\in J}\right)
\end{equation}
where $J \subset \{1, \ldots, L\}$ is the set of indices $j$ where $r_{j}>\epsilon$ and $\mathbbm{1}_{J}$ is the indicator vector for the subset $J$ (i.e. $(\mathbbm{1}_{J})_j = 1$ if $j \in J$ and $(\mathbbm{1}_{J})_j = 0$ if $j \notin J$). Conversely, given the code $\boldsymbol{\varepsilon}$ that contains error residual values, it is easy to compute an estimate of the original residual by constructing a vector whose zeros are determined by $\mathbbm{1}_{J}$ and whose nonzero entries are given by $\left(r_{j}\right)_{j\in J}$. We denote~this vector as $\mbox{residual}(\boldsymbol{\varepsilon})$.

\subsection{Training, Compression and Decompression}
After describing different components of our algorithms, we are now ready to integrate them. We assume that all the data mentioned in this section has been aligned and that all missing values have been imputed as described in Section~\ref{sec:missing_data}. For the training data $\mathbf{D}$, we also ensure that outliers were removed and that readings were normalized. Let $\sigma$ denote the standard deviation used in the normalization of the data.

We first learn optimal weights $\boldsymbol{\theta}$ for the autoencoder by minimizing the cost function $\Gamma_{\text{WAE}}(\boldsymbol{\theta},\mathbf{D})$ using the L-BFGS algorithm. This computationally-intensive process only occurs once at the start of our network deployment, and the parameters $\boldsymbol{\theta}, \sigma$ are distributed to the transmitters and receivers.

The algorithms for compressing and decompressing the sensor readings are outlined in Algorithms \ref{alg:compression} and \ref{alg:decompression}, respectively. For our experiments, we send the compressed signal $(\mathbf{y},\boldsymbol{\varepsilon},m)$ using floating point representation for the real numbers and binary string for the indicator vector $\mathbbm{1}_{J}$ in $\boldsymbol{\varepsilon}$. Note that all the steps have low computational complexity. Here, we also see why decoder complexity in algorithms (e.g., compressed sensing) impedes the provision of error bound guarantee because it is computationally expensive to compute the residual $\boldsymbol{\varepsilon}$. Clearly, an intruder who can receive the compressed data cannot retrieve the raw data without knowing the decoding weight matrix $\mathbf{W}_{dec}$ and bias vector $\mathbf{b}_{dec}$. This adds a free level of security to data aggregation in WSNs.

\begin{algorithm}
	\DontPrintSemicolon
	\KwIn{readings $\mathbf{x}$; parameters $\sigma, \mathbf{W}_{enc},\mathbf{b}_{enc},\mathbf{W}_{dec},\mathbf{b}_{dec}$}
	\KwOut{signal $\mathbf{y},\boldsymbol{\varepsilon},m$}
	\Begin{
	$m \leftarrow \text{mean}(\mathbf{x})$\;
	$\mathbf{d} \leftarrow \text{normalize}(\mathbf{x},\sigma)$\;
	$\mathbf{y} \leftarrow F (\mathbf{W}_{enc}\mathbf{d}+\mathbf{b}_{enc})$\;
	$\mathbf{\hat{d}} \leftarrow F(\mathbf{W}_{dec}\mathbf{y}+\mathbf{b}_{dec})$\;
	$\mathbf{p} \leftarrow \text{denormalize}(\mathbf{\hat{d}},m,\sigma)$\;
	$\boldsymbol{\varepsilon} \leftarrow  \text{residualCode}(\mathbf{x}-\mathbf{p}, \epsilon)$\;
	}
\caption{The online data compression \label{alg:compression}}
\end{algorithm}

\begin{algorithm}
	\DontPrintSemicolon
	\KwIn{signal $\mathbf{y},\boldsymbol{\varepsilon},m$; parameters $\sigma,\mathbf{W}_{dec},\mathbf{b}_{dec}$}
	\KwOut{reconstruction $\mathbf{\hat{x}}$}
	\Begin{
	$\mathbf{\hat{d}} \leftarrow F(\mathbf{W}_{dec}\mathbf{y}+\mathbf{b}_{dec})$\;
	$\mathbf{p} \leftarrow \text{denormalize}(\mathbf{\hat{d}},m,\sigma)$\;
        $\mathbf{r} \leftarrow \text{residual} (\boldsymbol{\varepsilon})$\;
	$\mathbf{\hat{x}} \leftarrow \mathbf{p} + \mathbf{r} $\;
	}
\caption{The online data decompression \label{alg:decompression}}
\end{algorithm}

\subsection{Time Complexity}
Our algorithm training is computationally expensive and should be run on a server. However, the data compression and decompression, as highlighted in Algorithms~\ref{alg:compression} and \ref{alg:decompression}, are lightweight. Both data compression and decompression has a linear time complexity of $\mathcal{O}\left(L\times K\right)$, where $L$ is the input data size, and $K$ is the compressed data size. This low computational complexity results in significant energy conservation as shown in Section~\ref{sec:energy_conservation}.

The next section presents simulation results to show the compression performance and energy conservation of the proposed algorithm.

\section{Experimental Results }\label{sec:experimental}

We evaluate the performance of the proposed algorithm using data from actual sensor test beds. Our datasets are divided into 10 random folds for training and testing (i.e., the cross-validation method~\cite{kohavi1995study} with 10 folds). In each cross-validation step, the system is trained using 9 folds and tested using the last fold. Our implementation adopts the limited memory Broyden\textendash{}Fletcher\textendash{}Goldfarb\textendash{}Shanno (L-BFGS) algorithm~\cite{byrd1995limited} to tune the AE's weights during the learning stage.

\subsection{Datasets and Performance Metrics}
We evaluate our solution using the following meteorological datasets:
\begin{itemize}
\item Grand-St-Bernard deployment~\cite{sensorscope2014}: We use data from 23 sensors that collect surface temperature readings between Switzerland and Italy at an elevation of 2.3km. This dataset contains readings ranging from $-32^{\circ}\text{C}$ to $48^{\circ}\text{C}$, though observations suggest that the maximum and minimum values are most likely from malfunctioning sensor nodes. After removing outliers, the dataset still contains many interesting nontrivial features.
\item LUCE deployment~\cite{sensorscope2014}: After data preparation, the used dataset contains relative humidity measurements from 90 sensors, each with around 160k readings. This dataset is an example of high resolution spatial-temporal data that is collected by WSNs to monitor an area with widely varying data characteristics.
\end{itemize}

To measure the extent that the data is being compressed, we use the following metrics: 
\begin{itemize}
\item \emph{Compression ratio} (CR): This metric calculates the reduction in transmitted data size due to compression which is defined as follows:
\begin{equation}
\text{CR}(\mathbf{x},\mathbf{\hat{x}})=\left(\frac{B(\mathbf{\hat{x}})}{B(\mathbf{x})}\right)\times100
\end{equation}
where $B(\mathbf{\hat{x}})$ and $B(\mathbf{x})$ are the numbers of bits used to represent the transmitted and the original data, respectively.
\item \emph{Root mean squared error} (RMSE): RMSE measures the loss of data precision due to compression algorithms, i.e., compression error. An RMSE of~$0$ means that WSN data can be fully reconstructed without error. RMSE is defined as follows:
\begin{equation}
\text{RMSE}(\mathbf{x},\mathbf{\hat{x}})=\sqrt{\frac{1}{L}\sum_{i=1}^{L}(x_{i}-\hat{x}_{i})^{2}}.
\end{equation}
\item \emph{Coefficient of determination} (usually denoted by $R^{2}$ in statistics): This defines the proportion of variance of the original data that is reconstructed from the compressed data. An $R^{2}$ of $0.4$ means that 40\% of $\mathbf{x}$ is reconstructed in $\mathbf{\hat{x}}$. This metric is calculated as follows:
\begin{equation}
R^{2}(\mathbf{x},\mathbf{\hat{x}})=1.0-\frac{\sum_{i=1}^{L}(x_{i}-\hat{x}_{i})^{2}}{\sum_{i=1}^{L}(x_{i}-\bar{x})^{2}},
\end{equation}
where $\bar{x}=\frac{1}{L}\sum_{i=1}^{L}x_{i}$. The data is fully reconstructed if $R^{2}$ is equal to $1.0$.
\end{itemize}

The CR value determines the compression efficiency, while the RMSE and $R^{2}$ values define the reconstruction fidelity.

\subsection{Test Example}
\begin{figure*}
\begin{centering}
\subfloat[{\label{fig:overview_example_1}\small{}}]
{\begin{centering}
\includegraphics[width=0.6\columnwidth,trim=0.5cm 1cm 0.5cm 1cm]{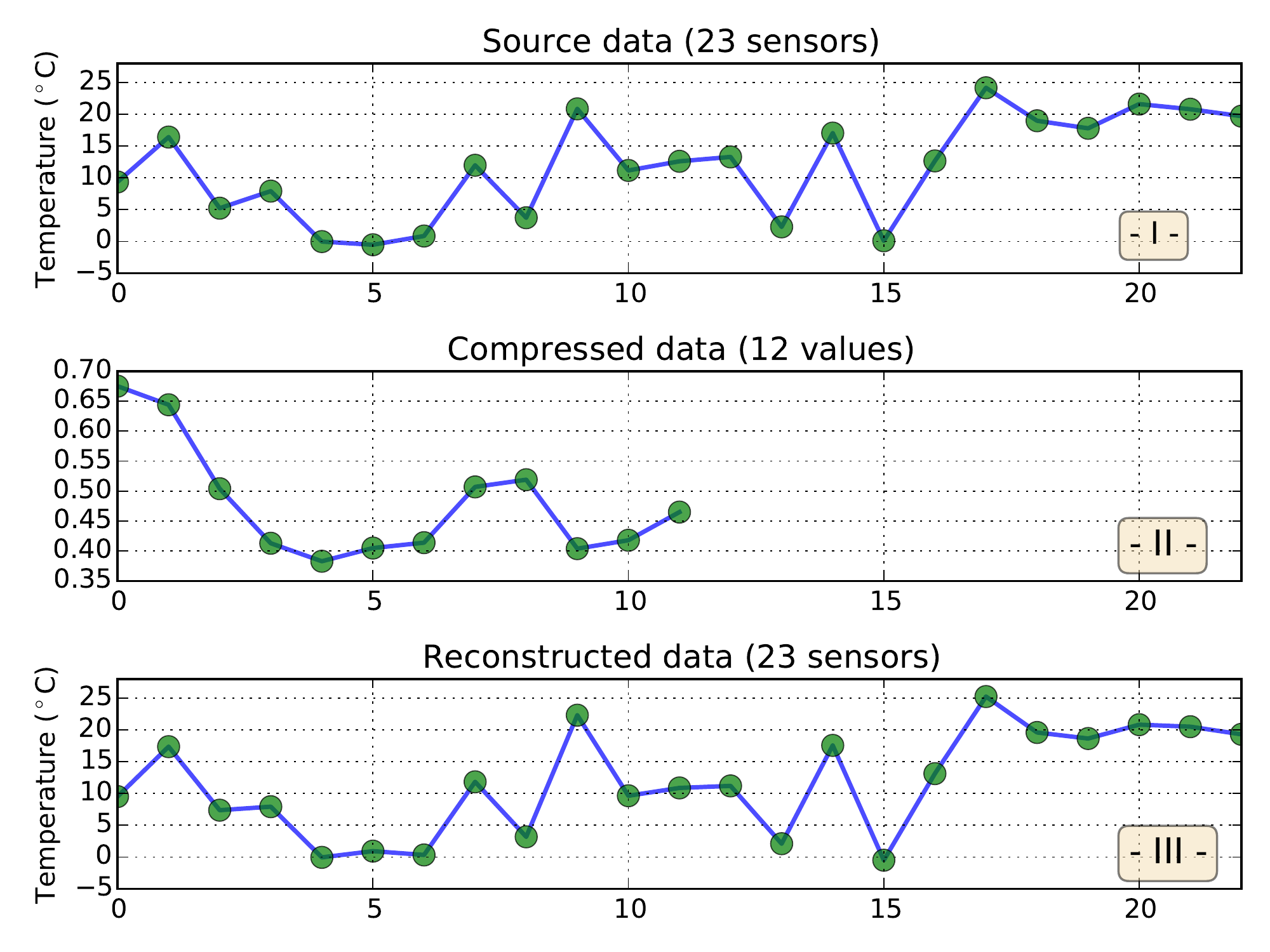}
\par\end{centering}}
\enskip{}
\subfloat[{\label{fig:overview_example_2}\small{}}]
{\begin{centering}
\includegraphics[width=0.6\columnwidth,trim=1cm 2cm 2cm 2cm]{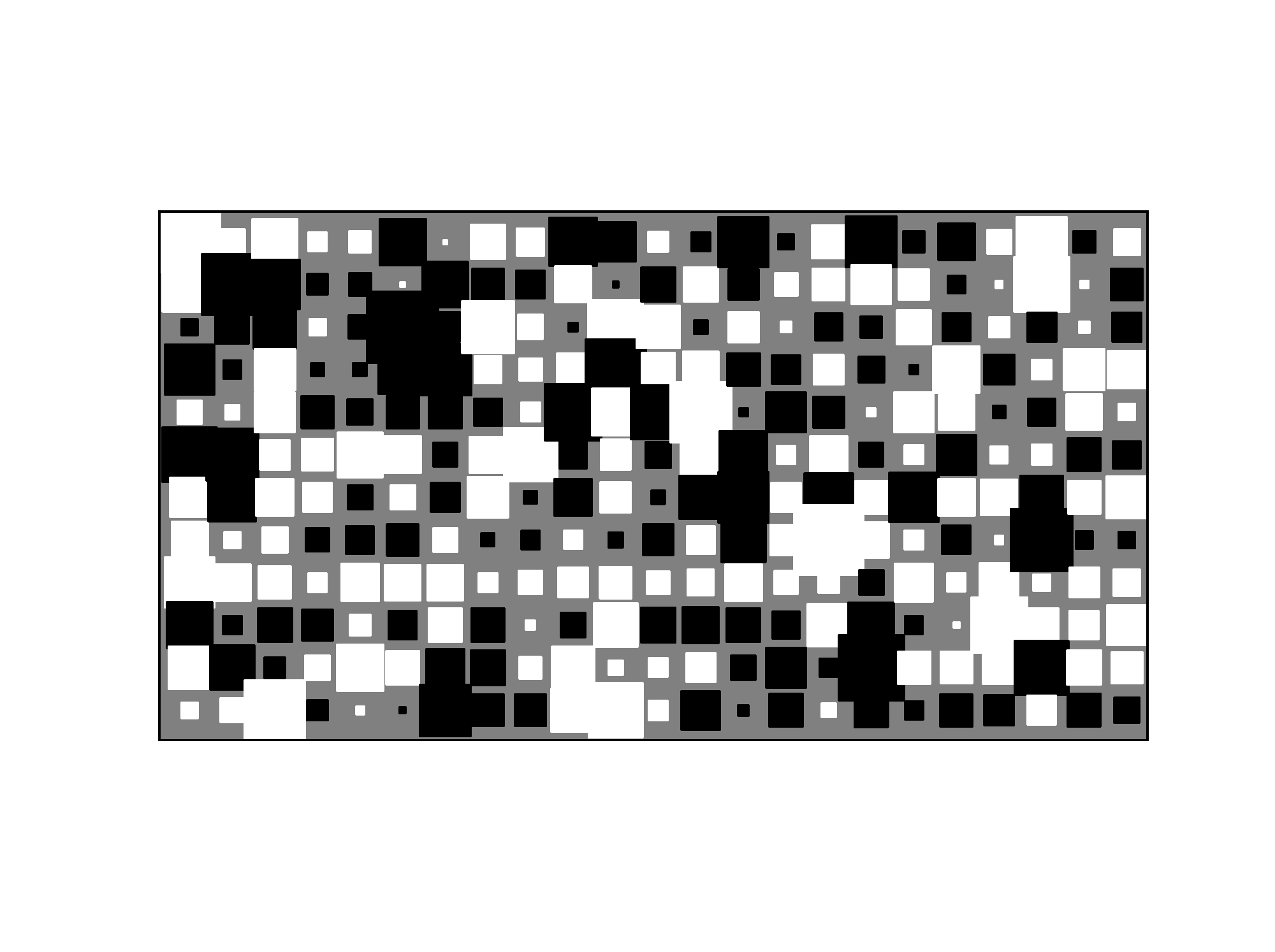}
\par\end{centering}}
\enskip{}
\subfloat[{\label{fig:overview_example_3}\small{}}]
{\begin{centering}
\includegraphics[width=0.7\columnwidth,trim=2cm 0.5cm 2cm 0.3cm]{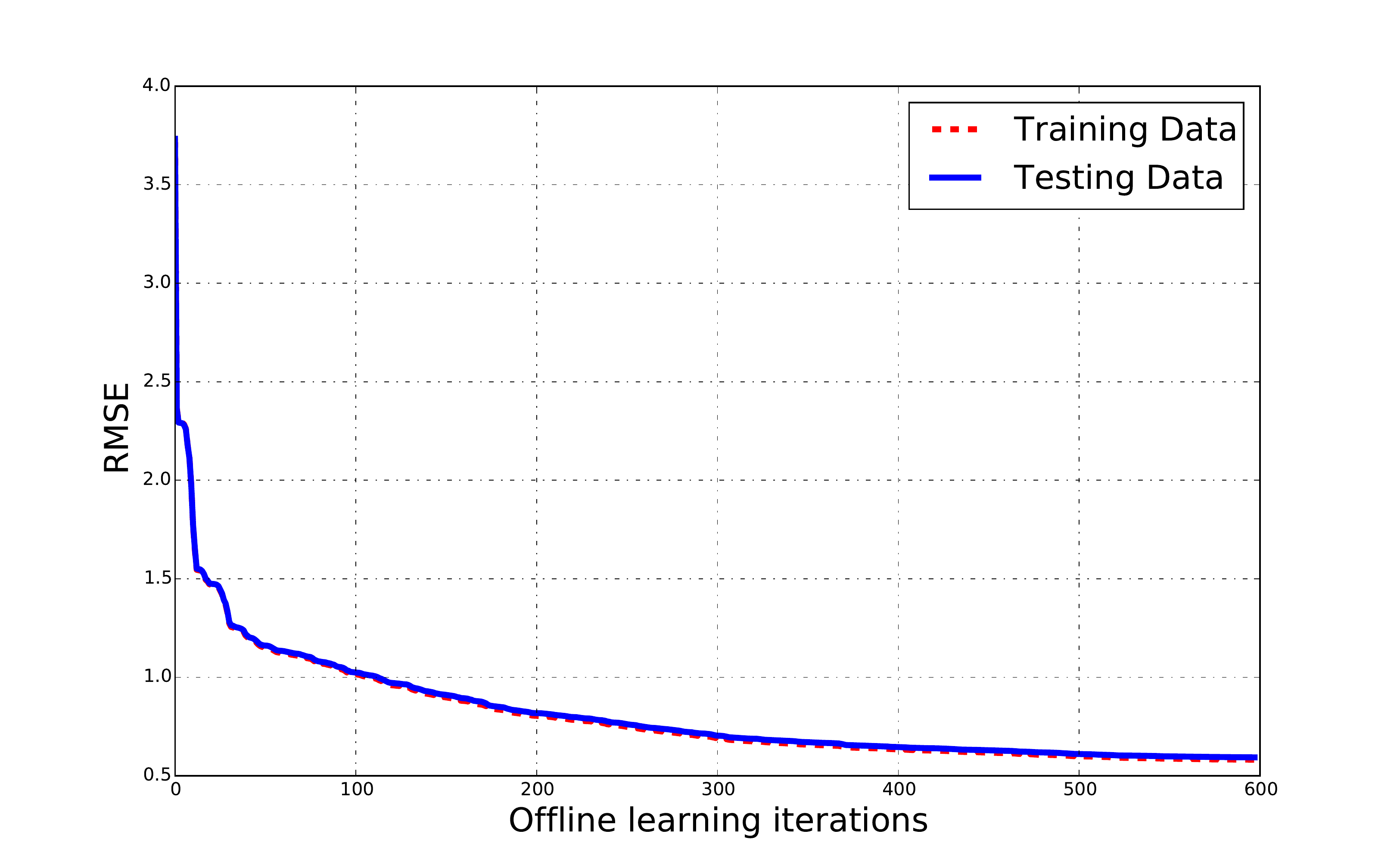}
\par\end{centering}}
\par\end{centering}

\caption{\label{fig:overview_example}An illustrative data compression example of surface temperature readings in a WSN containing 23 sensors: (a) an example of a data vector compression and recovery, (b) a Hinton diagram to illustrate the learned weights of the encoder, and (c) the offline learning curve over training and testing datasets.}
\end{figure*}

Figure~\ref{fig:overview_example} provides an example of the quantities computed in the proposed compression solution. The network is trained using $28k$ records of historical dataset from the Grand-St-Bernard dataset. Figure~\ref{fig:overview_example_1} gives an example of data compression, transmission and recovery process using the AE's network. The input signal (I) is collected from the network's nodes, such that each node contributes one reading every two minutes. This input signal excites the network and a compressed signal (II) is generated and transmitted to the receiver(s) using any general routing protocol. The output signal (III) is recovered at the receiver that represents an efficient approximation of the input signal. Figure~\ref{fig:overview_example_2} shows a Hinton diagram of the learned encoding weight. The size of the squares represents the filter's magnitudes and the color represent the sign (white for negative and black for positive values). Each column shows the receptive field of each node in the other nodes. The node's receptive fields are automatically extracted to represent the spatial correlation among neighbor nodes. Figure~\ref{fig:overview_example_3} shows the RMSE over learning iterations for the training and testing datasets. The training RMSE is very high at the initial iterations but decreases with learning iterations.

\subsection{Baselines}

In this section, a simulation study of the data compression is given. This includes two main validation scenarios. Firstly, the algorithm performance without error bound guarantee is tested under spatial compression scenario using the Grand-St-Bernard deployment. This scenario is designed to test the compression ratio and  reconstruction error of the proposed method against a set of conventional methods that do not provide any error bound guarantee. Secondly, a temporal compression scenario is formulated with an error bound guarantee using the LUCE deployment data. This temporal scenario tests the proposed method against the well-known LTC method~\cite{schoellhammer2004lightweight} which provides an error bound guarantee.

\subsubsection{AE Models}

\begin{figure}
\begin{centering}
\includegraphics[width=0.75\columnwidth,trim=1cm 0.8cm 1cm 1cm]{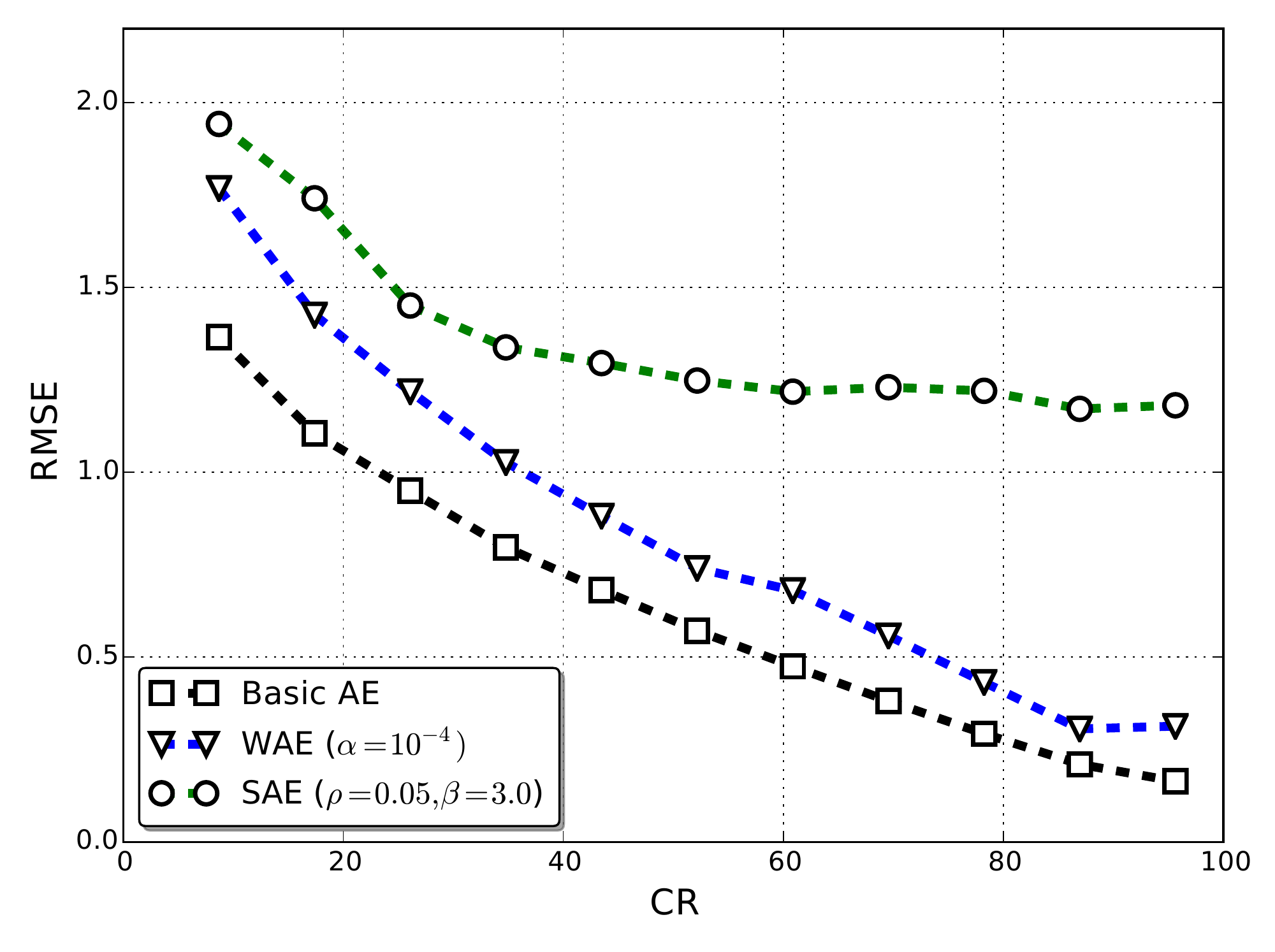}
\par\end{centering}
\caption{\label{fig:ae_variants}Compression error (RMSE) achieved by several AE models under varied compression ratio (CR).}
\end{figure}

As shown in Figure~\ref{fig:ae_variants}, using the basic AE provides the best performance over the other AE's variants. Even though WAE and SAE are useful for classification-related tasks to avoid overfitting, we find that they degrade the AE's reconstruction performance (i.e., RMSE). This is justified as for the compression problem, the hidden layer size is less than the input size, and hence the model is less affected by the overfitting problem. However, in feature extraction and classification problems, the hidden layer could be larger than the input size and the overfitting effects are more apparent. In these cases, regulations using WAE and SAE become more important. To tune the AE's hyper-parameters, the authors of~\cite{coates2011analysis} describe the prohibitive task of running the cross-validation procedure using several choice of parameters. To automate this process, we employed the common strategy in machine learning by using the grid search method for model selection. Initially, this starts by specifying a list of reasonable values of each hyperparameter. Then, the algorithm is evaluated over the elements of the cross product set of the lists. In summary, the parameters that achieve the best performance on the cross-validation estimator will be chosen for real time deployment. It is important to note that the grid search method becomes ineffective for a large number of hyperparameters as the lists' product increases dramatically~\cite{do2007efficient}. However, we only have two hyperparameters in the sparse AE case.

\subsubsection{Spatial Compression}

Without the error bound guarantee, we use the Grand-St-Bernard dataset to test the spatial compression capabilities of the proposed algorithm. The 23 sensors are assumed to be synchronized to transmit their data samples to a gateway. The gateway will spatially compress the data before sending it to the BS over a backhaul link. This data compression is a challenging task due to the non-uniform data distribution through different sensor nodes.

\begin{figure*}
\begin{centering}
\subfloat[{\label{fig:experimental_validation_stbernard_1}}]
{\begin{centering}
\includegraphics[width=0.75\columnwidth,trim=1cm 0.5cm 1cm 1cm]{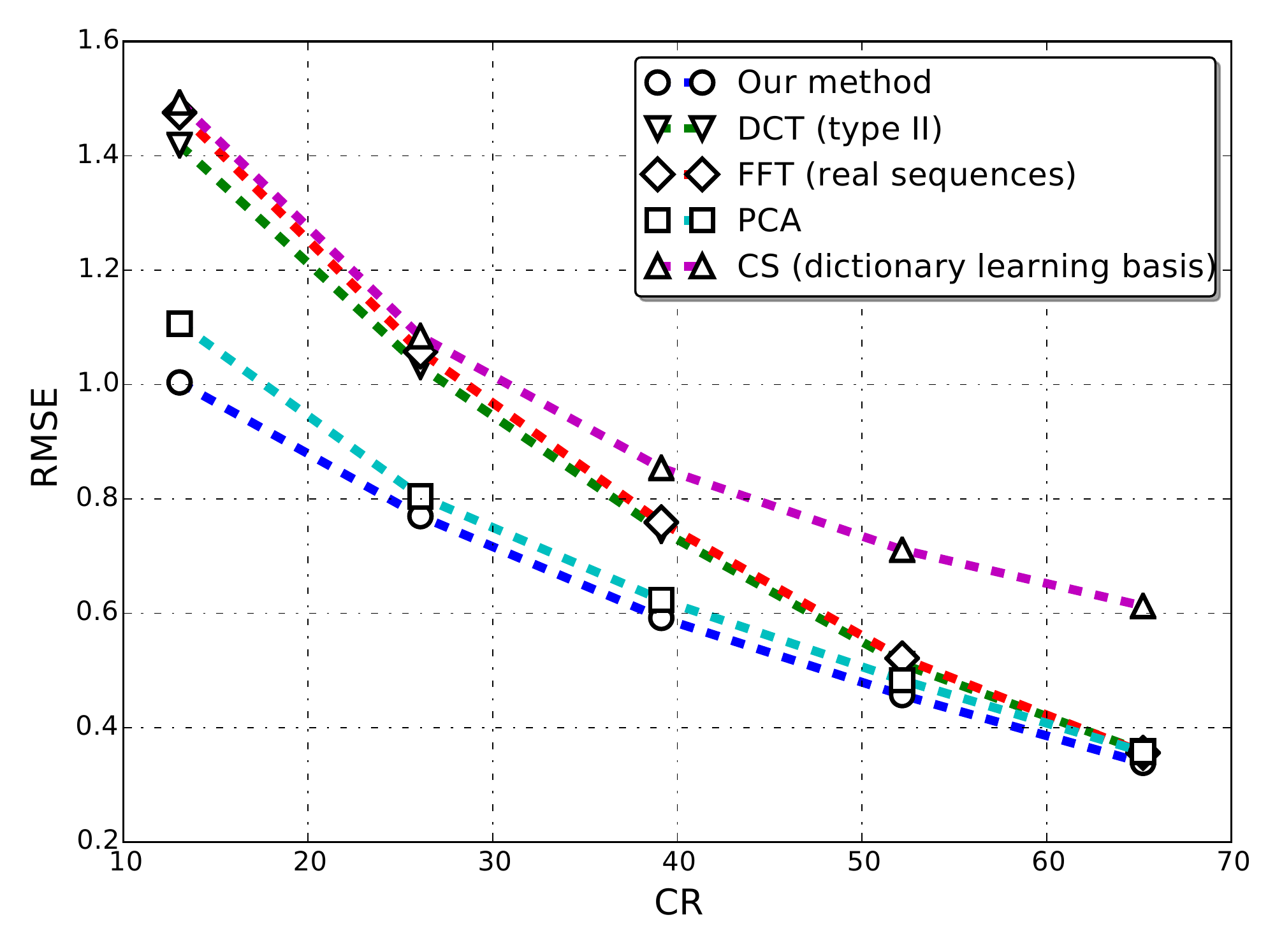}
\par\end{centering}
}\enskip{}\enskip{}
\subfloat[{\label{fig:experimental_validation_stbernard_2}}]{\begin{centering}
\includegraphics[width=0.75\columnwidth,trim=1cm 0.5cm 1cm 1cm]{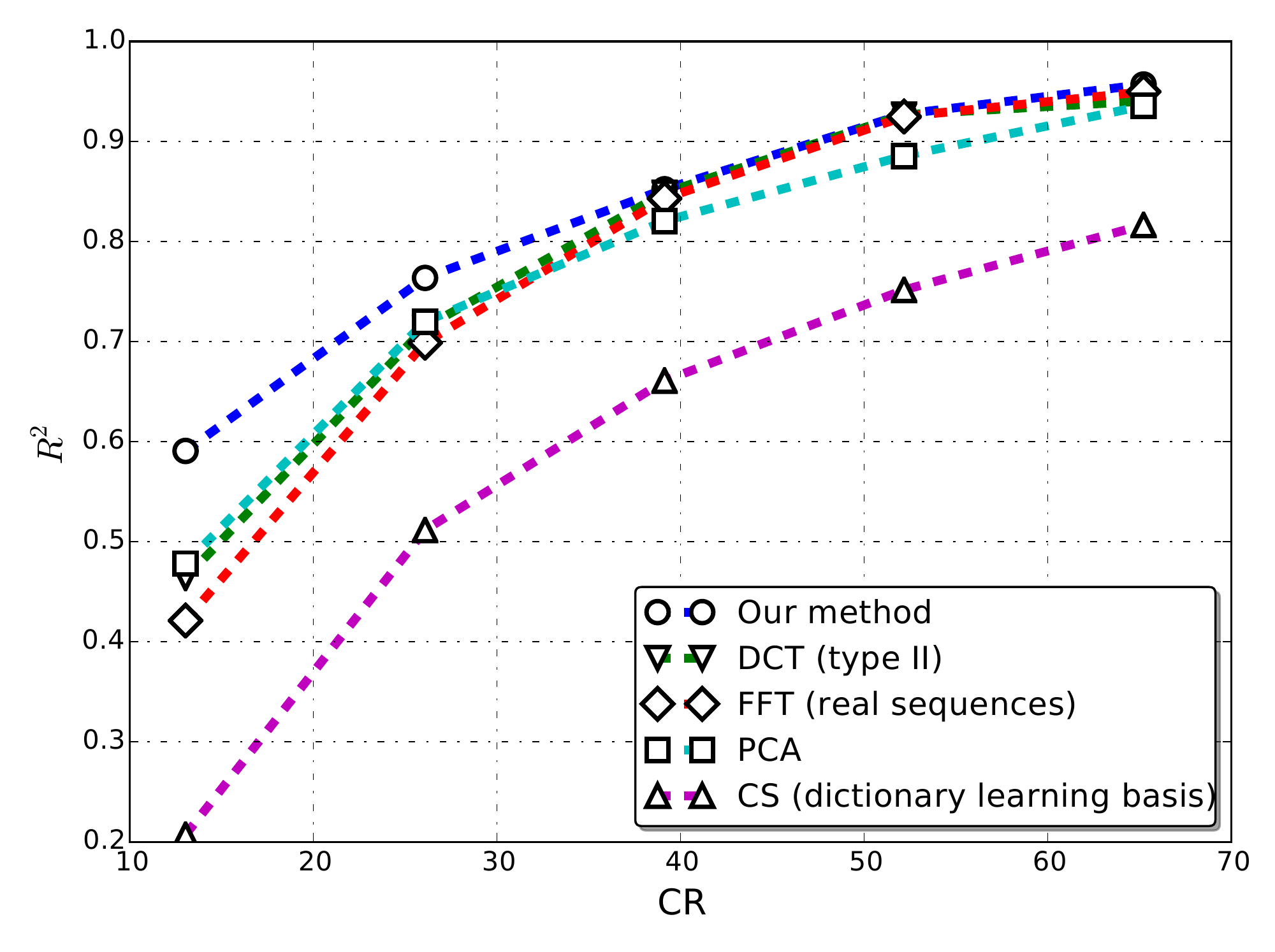}
\par\end{centering}}
\par\end{centering}

\caption{Reconstruction fidelity of a spatial compression scenario without error bound guarantee. This shows the performance of the proposed method and conventional methods on the Grand-St-Bernard dataset. (a)~Compression error (RMSE) under different values of compression ratio (CR). (b)~Coefficient of determination ($R^{2}$) under different values of compression ratio (CR).}\label{fig:experimental_validation_stbernard}
\end{figure*}

Figures~\ref{fig:experimental_validation_stbernard_1} and \ref{fig:experimental_validation_stbernard_2} show that the proposed method outperforms other conventional WSN data compression methods such as PCA, DCT, FFT, and CS. These conventional methods are the main basis for most existing methods for WSN data compression~\cite{razzaque2013compression,zordan2012compress}. Our implementation of these conventional methods is based on the scikit-learn library~\cite{pedregosa2011scikit}. CS samples data at a low rate than Shannon Nyquist sampling rate. Specifically, an input signal $\textbf{x}\in\mathbb{R}^{L}$, $L\in\left\{N,M\right\}$ is represented as $\mathbf{x}=\mathbf{\Psi}\mathbf{s}$, where $\mathbf{s}$ is the sparse representation of the signal with $\alpha$ nonzero values (called $\alpha$-sparse), and $\mathbf{\Psi}\in\mathbb{R}^{L \times L}$ is the basis dictionary. We have used online dictionary learning~\cite{mairal2009online} to find $\mathbf{\Psi}$. Other limitations of CS as a WSN data compression method have been discussed in Section~\ref{sec:CS_limitations}.

\begin{figure*}
\begin{centering}
\includegraphics[width=0.75\paperwidth,trim=1cm 0.5cm 1cm 0cm]{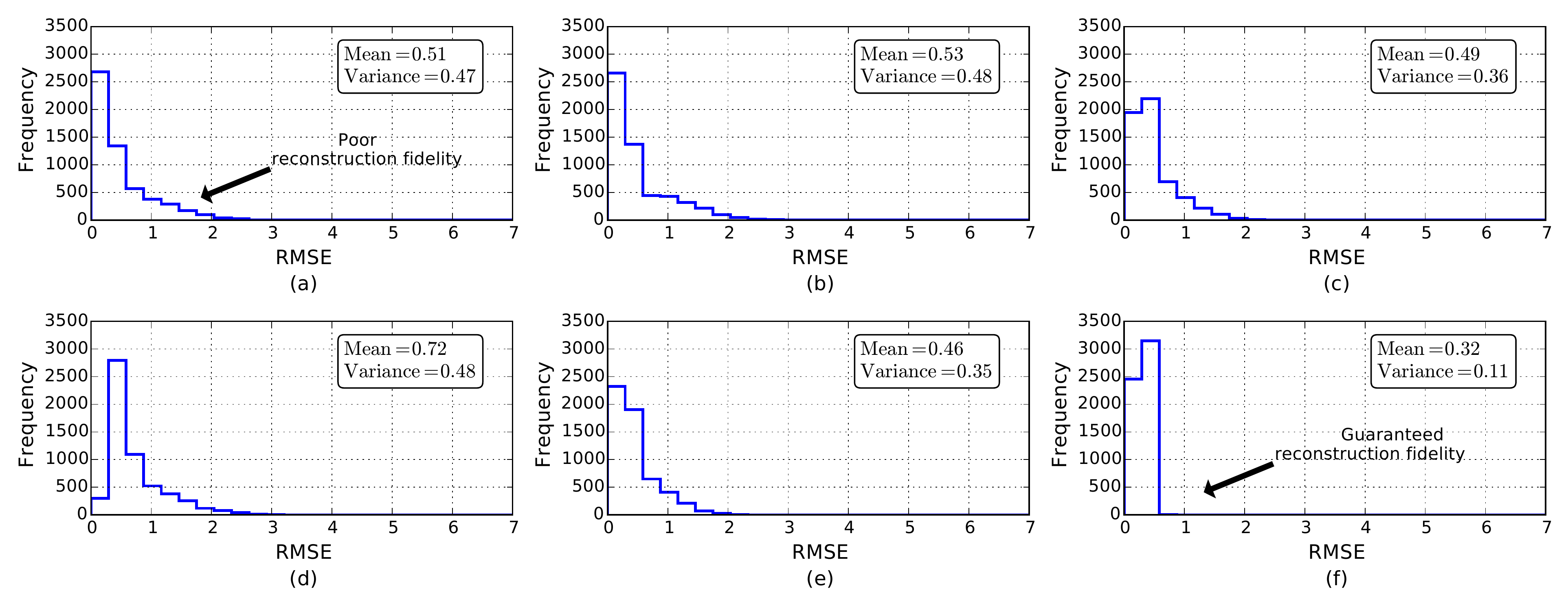}
\par\end{centering}

\caption{Error bars of a spatial compression scenario on the Grand-St-Bernard dataset. (a)-(e)~WSN data compression methods without error bound guarantee achieving a compression ratio of $\text{CR}=52.2\%$. (a)~DCT, (b)~FFT, (c)~PCA, (d)~CS with dictionary learning basis, and (e) the proposed method. (f)~The proposed method with error bound guarantee (error bound $\epsilon=1.0$) resulting in a compression ratio of $\text{CR}=60.6\%$.}\label{fig:err_bar}
\end{figure*}

Based on Figure~\ref{fig:err_bar}, we observe an important result. The average RMSE value can be misleading as WSN data compression methods without error bound guarantee can produce poor reconstruction fidelity at some time instants. Most traditional lossy data compression algorithms in WSNs lack an error bound guarantee mechanism due to the high computational demand of data decompression and reconstruction~\cite{razzaque2013compression}. Our proposed method overcomes this limitation by using the error bound mechanism proposed in Section~\ref{sec:error_bound_mechanism}. The proposed method with error bound of $\epsilon=1.0$ gives a good compression ratio of $\text{CR}=60.6\%$.

\subsubsection{Temporal Compression}
In the following, we compare the proposed method with the LTC method in a temporal compression scenario with error bound guarantee. We choose the LTC algorithm for bench-marking which as (1)~LTC is one of the rare WSN data compression methods with error bound guarantee, and (2)~several comparative studies (e.g.,~\cite{zordan2012compress}) discussed the efficiency of the LTC algorithm over other methods in temporal compression.

Using the LUCE deployment, the temporal compression scenario is formulated such that each sensor compresses its data locally before sending it using multihop transmissions to the BS. Each sensor is assumed to sample at a rate of 1 sample every 2 minutes. Therefore, 720 samples are collected each day and sent as one compressed chunk.

\begin{figure*}
\begin{centering}
\subfloat[{\label{fig:experimental_validation_luce_1}\small{}}]
{\begin{centering}
\includegraphics[width=0.75\columnwidth,trim=1cm 0.5cm 1cm 1cm]{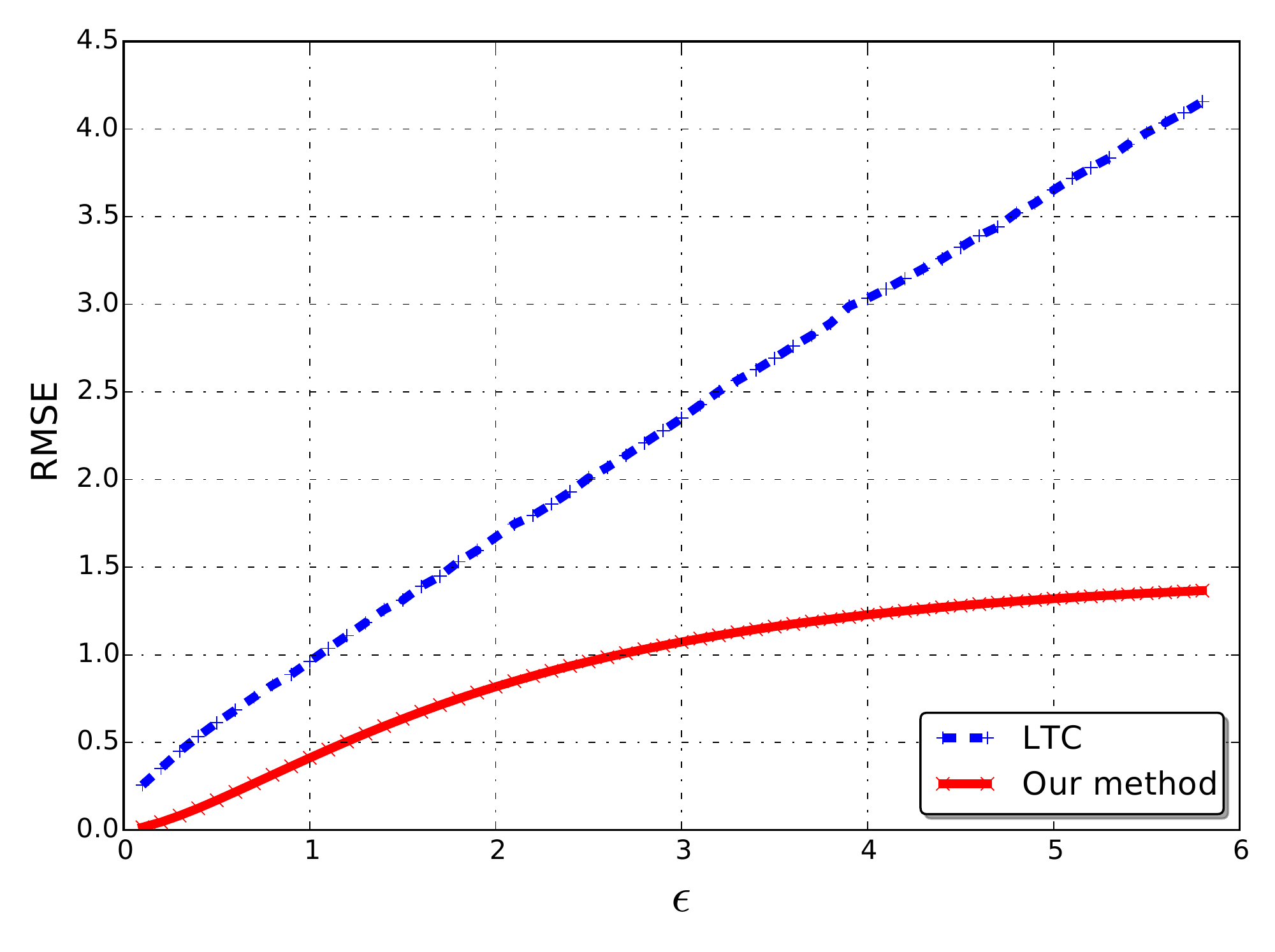}
\par\end{centering}
}\enskip{}\enskip{}
\subfloat[{\label{fig:experimental_validation_luce_2}\small{}}]
{\begin{centering}
\includegraphics[width=0.75\columnwidth,trim=1cm 0.5cm 1cm 1cm]{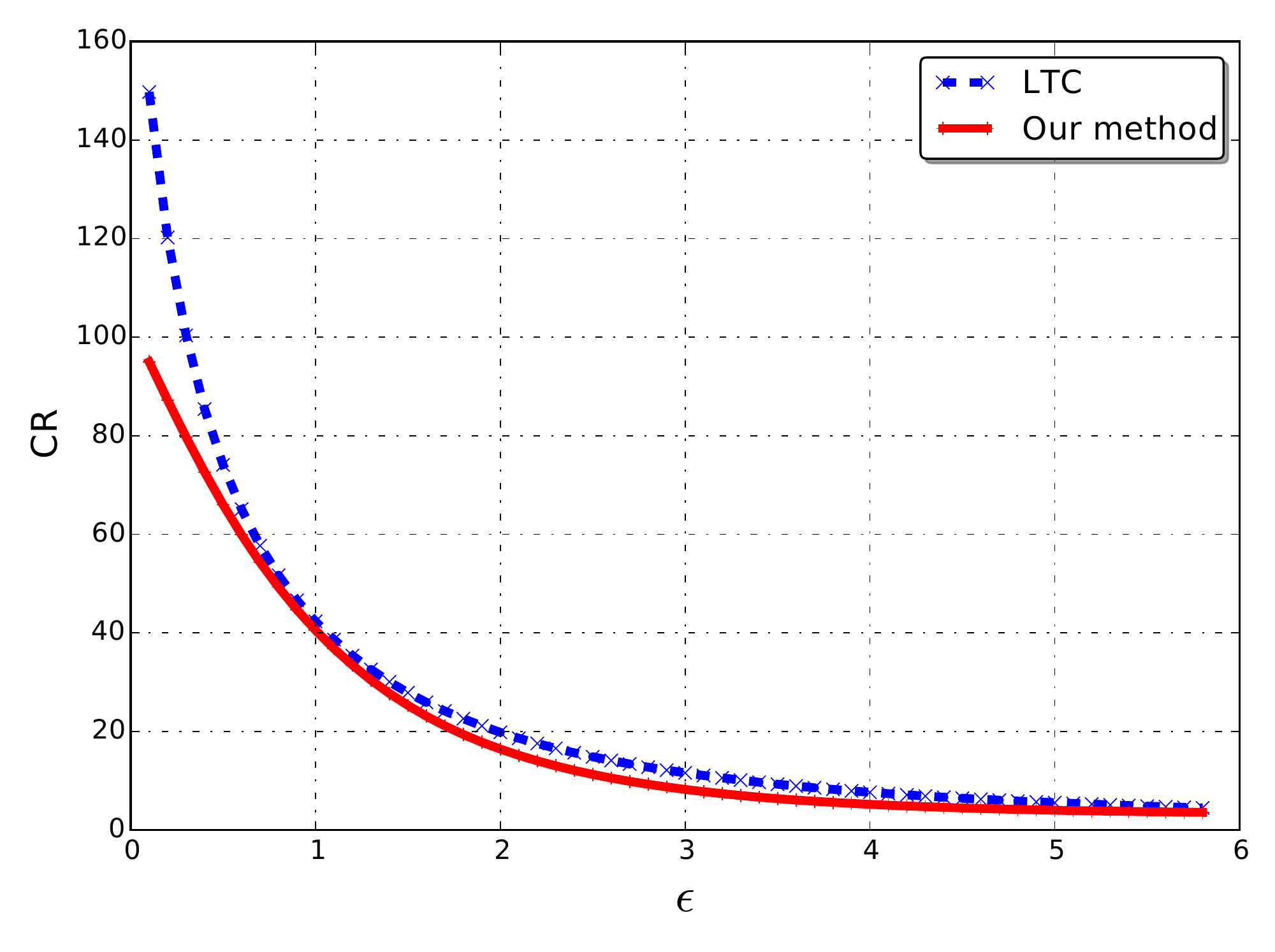}
\par\end{centering}
}\par\end{centering}
\caption{Analyzing a temporal compression scenario with error bound guarantee using the LUCE deployment dataset. The AE's input vector size is 720 and the hidden layer size is 20, so the achievable compression ratio is at most 97.22\%. (a)~Compression error (RMSE) under different values of error bound ($\epsilon$), and (b)~compression ratio (CR) under different values of error bound ($\epsilon$).}\label{fig:experimental_validation_luce}
\end{figure*}

Figure \ref{fig:experimental_validation_luce} provides the analysis of the data compression with an error bound constraint. This shows that the proposed algorithm outperforms the LTC in both RMSE (Figure \ref{fig:experimental_validation_luce_1}) and compression ratio (Figure \ref{fig:experimental_validation_luce_2}). Even though the high resolution dataset of the LUCE deployment is very suitable for the LTC method as the data changes slowly between subsequent samples, the compression efficiency of the proposed algorithm is still superior. We note that LTC performs as good as AE for large error bounds, but is unable to keep the same efficiency when the error bound is small.

\subsection{Energy Conservation by Data Compression}\label{sec:energy_conservation}

In this section, we consider the energy conservation of the proposed method. Traditional data compression schemes from information and coding theory cannot be directly applied to a resource limited framework like WSNs as they are designed to optimize storage rather than energy consumption~\cite{razzaque2013compression}. Therefore, special attention should be provided to the computational burdens of any compression algorithm designed for WSNs. Otherwise, the energy consumed during CPU operations of complex algorithms might exceed the energy consumed due to sending less data over the RF module.

Again, suppose that the length of the original data vector is $L$ and the length of the compressed data representation is $K$. We adapt the complexity analysis used in~\cite{zordan2012compress} while considering the power consumption for receiving the data which is extremely important in multihop data aggregation.
\begin{itemize}
\item We consider the mixed-signal MSP430 micro-controller~\cite{bierl2000msp430} that is powered with a 16-bit CPU developed to support resource limited systems. The supply voltage is $V_{CC}=3.3V$, the clock rate is $F_{CLK}=3.3 MHz$, and the current consumption of the complete MSP430 system during the active mode is $I_{MSP430}=1.85mA$ (see Section~5.3.4 of~\cite{bierl2000msp430}). Hence, the power consumption of the MSP430 micro-controller per clock cycle is
\[
E_{CLK}=\frac{V_{CC}\times I_{MSP430}}{F_{CLK}}=1.85nJ.
\]
The exponential function can be calculated using two hyperbolic functions as $\exp(\upsilon)=sinh(\upsilon)+cosh(\upsilon)$, which requires (without hardware multiplier) 52000 CPU cycles to achieve more than 6 digits of precision. This derivation is based on the Taylor series expansion. Therefore, it is important to select the number of Taylor iterations of the exponential function calculation to satisfy the precision requirements of the application~\cite{braverman2006computing}. The CPU cycle specifications of the basic operations are given in Table~\ref{tab:cpu_operations}.

\item For the transmission unit, we consider the 9XTend RF module~\cite{manual20059xtend} that operates in the $902-928MHz$ frequency band with an effective data rate of $R_{XTend}=9,600bps$ and a spread technology of frequency-hopping spread spectrum (FHSS). This module's transmission range is of up to $0.9km$ in urban areas, and up to $22km$ for ideal outdoor line-of-sight transmissions. These transmission ranges make XTend module suitable for data transmission over a backhaul link. The current consumption during the data transmission and reception are $I_{TX}=600mA$ and $I_{RX}=80mA$, respectively%
\footnote{Some studies ignore the power consumption of data reception. However, this metric is important in multihop transmission which is performed by regular sensor nodes with limited energy budget. %
}. The current flow during the idle mode is near $1mA$, and hence it is ignored in our analysis. The supply voltage is set at $V_{CC}=3.3 V$. Then, the consumed energy for transmitting and receiving one bit of data is
\[
S_{bit}=\frac{V_{CC}\times I_{TX}+V_{CC}\times I_{RX}}{R_{CC2420}}=233.75\mu J.
\]

\end{itemize}

\begin{table}
\begin{centering}
\begin{tabular}{|l|c|}
\hline 
\textbf{Operation} & \textbf{\# of CPU Cycles (FLOAT)}\\
\hline 
\hline 
Addition & 184\tabularnewline
\hline 
Subtraction & 177\tabularnewline
\hline 
Multiplication & 395\tabularnewline
\hline 
Division & 405\tabularnewline
\hline 
Comparison  & 37\tabularnewline
\hline 
$\exp\left(\cdot\right)$& 52000\tabularnewline
\hline 
\end{tabular}
\par\end{centering}

\caption{\label{tab:cpu_operations}CPU clock cycles for the mixed-signal MSP430 micro-controller~\cite{bierl2000msp430}.}
\end{table}

Therefore, the energy consumed by the network to transmit one bit and receive it at the next hop (one hop transmission) over the transceiver unit is approximately equal to the energy used by the micro-controller in $125,945$ CPU clock cycles\footnote{Larger ratios of the transmission-CPU energy consumptions are even given in other studies, see~\cite{sadler2006data} as an example, which is based on the hardware set and the CPU energy saving modes. These larger ratios result in more energy savings when using data compression algorithms.%
}. Using these design components, we formulate the computational complexity (in number of clock cycles) for compressing the input vector using the AE network $C_{AE}\left(L,K\right)$ as:

\begin{multline}
C_{AE}\left(L,K\right)=\underbrace{(184+405+177+2\times37)L}_{\text{Data normalization}}+\\
\underbrace{(395L+2\times184L)K}_{\mathbf{W}_{enc}\mathbf{x}+\mathbf{b}_{enc}}+\underbrace{(184+405+52000)K}_{\text{The sigmoid function}}.\label{eq:ae_complixity}
\end{multline}
Finally, using (\ref{eq:ae_complixity}), we find that the energy consumed to transmit the data with compression can be formulated as:

\begin{equation}
E_{AE}\left(L,K\right)=\underbrace{E_{CLK}\times C_{AE}\left(L,K\right)}_{\text{CPU cost}}+\underbrace{32\text{bits}\times K\times S_{bit}}_{\text{Transmission cost}}.\label{eq:compressing_energy}
\end{equation}

The first term refers to the energy consumed to compress the data (i.e. during the CPU computations), and the second term considers the energy consumed at the transmission unit to send the compressed bits. Note that we consider a 32 bit float representation of the sensor readings. Clearly, to achieve energy savings, the energy consumption using data compression scheme must be significantly less than that of the transmission of the raw data, more formally:

\begin{equation}
E_{AE}\left(L,K\right)\ll32bits\times L\times S_{bit}.\label{eq:compressing_efficiency}
\end{equation}

\begin{figure*}
\begin{centering}
\subfloat[{\label{fig:energy_consumption_1}}]
{\begin{centering}
\includegraphics[width=0.75\columnwidth,trim=1cm 0.3cm 0.5cm 1cm]{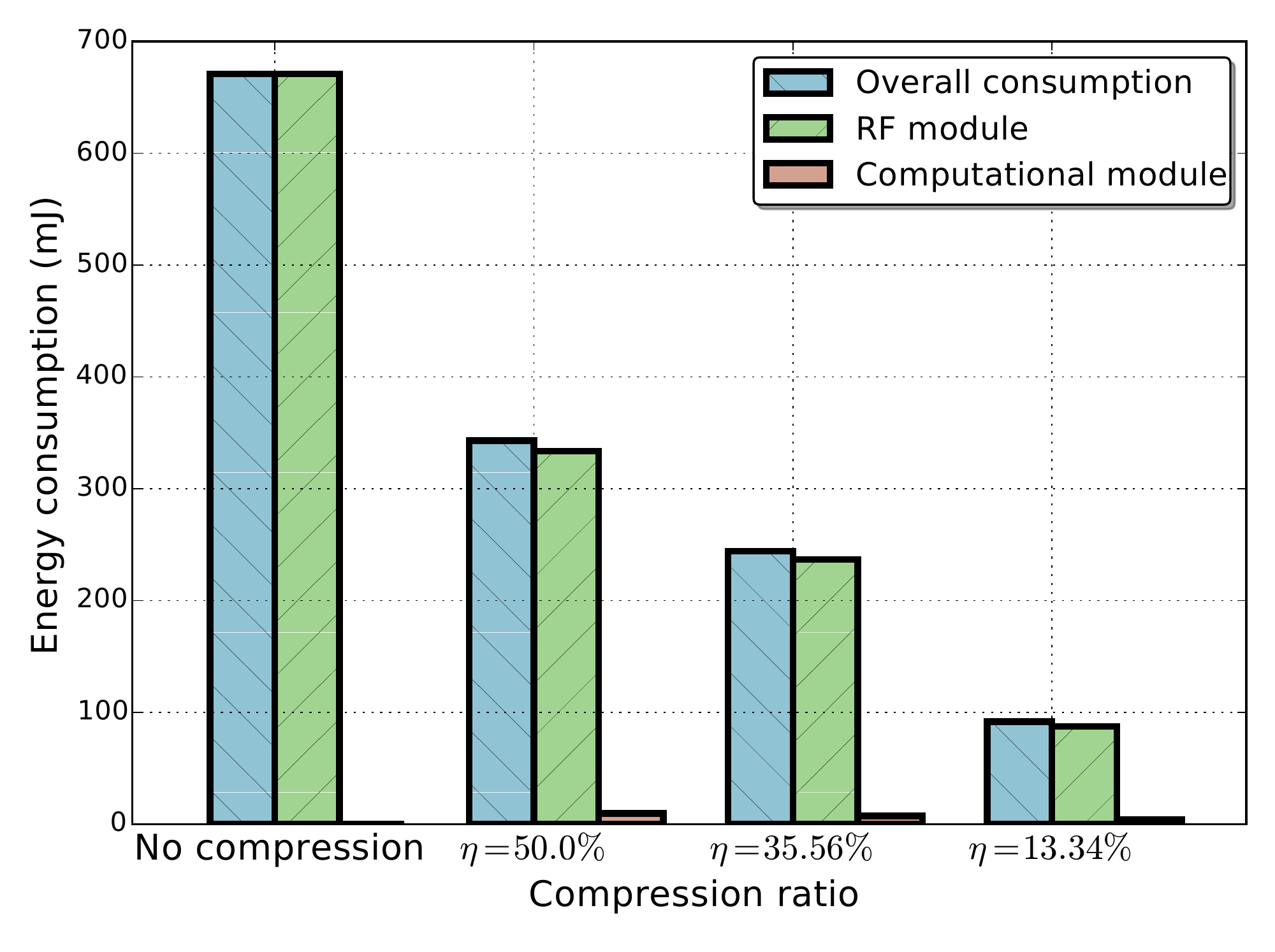}
\par\end{centering}}\enskip{}
\subfloat[{\label{fig:energy_consumption_2}}]
{\begin{centering}
\includegraphics[width=0.75\columnwidth,trim=0.5cm 0.3cm 1cm 1cm]{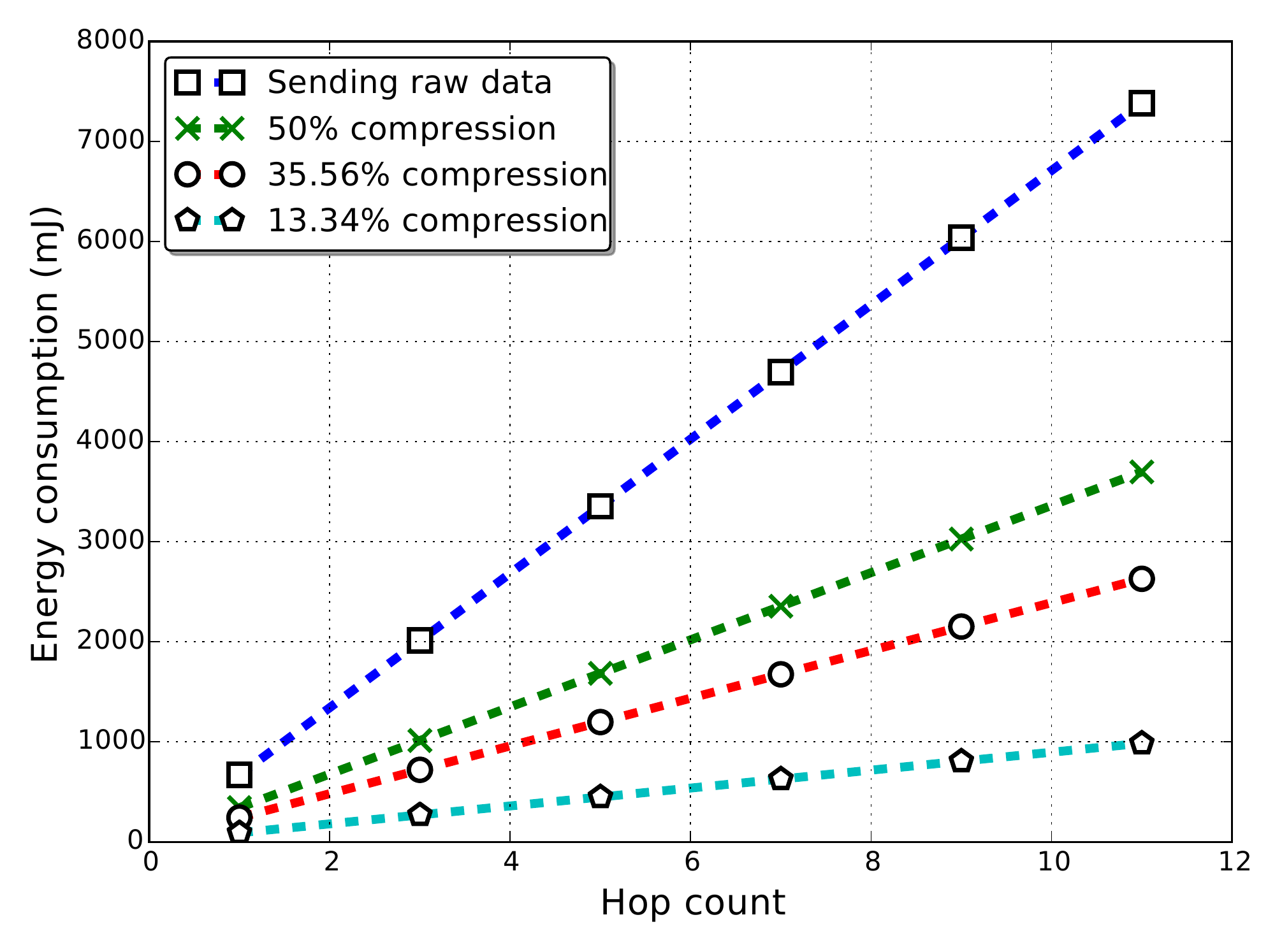}
\par\end{centering}
}\par\end{centering}
\caption{\label{fig:energy_consumption}{\small{}Energy conservation by data compression, assuming raw data of 90 sensors that is spatially compressed. (a)~Energy consumption at different compression ratios. (b)~Energy consumption on multihop transmissions with different hop counts.}}
\end{figure*}

These results are illustrated in Figure~\ref{fig:energy_consumption} under different compression ratios and multihop transmissions. Specifically, Figure~\ref{fig:energy_consumption_1} shows the energy conservation by data compression at different compression ratios. Figure~\ref{fig:energy_consumption_2} shows the increased energy conservation by data compression for multihop data transmission where the forwarding nodes are typical sensor nodes with energy-limited budgets. For example, a CR of 35.56\%  in 5-multihop transmissions reduces the energy consumption by 2.8 folds as compared to raw data transmission. A similar result can be drawn for reliable networks in which several copies of the same packet is transmitted to ensure a packet delivery ratio.

\section{Conclusion}\label{sec:conclusion}
Instead of using computationally expensive transformations on raw data or introducing strong assumptions on data statistical models, we have proposed an adaptive data compression with feature extraction technique using AEs. Our solution exploits spatio-temporal correlations in the training data to generate a low dimensional representation of the raw data, thus significantly prolonging the lifespan of data aggregation and funneling systems. Moreover, the algorithm can optionally be adjusted to support error bound guarantee.

Recent sensor networks often monitor a variety of modalities such as temperature, humidity and illuminance. However, designing a compression algorithm for multimodal data is much more challenging than the single modal situation~\cite{Srisooksai201237}. To study fundamental issues and design tradeoffs, we ignore the case of multimodal data in this paper and keep it for a future work. We will also study the use of sparse over-complete representations for data compression in WSNs (i.e. when the hidden layer size is larger than the input size). Last but not least, we will explore how to integrate the presence of missing values into our autoencoder, rather than using a naive method for missing data imputation.

\section*{Acknowledgment}
We thank Puay Kai Poh for valuable discussions in the early stages of the study.

\bibliographystyle{IEEEtran}
\bibliography{references}

\vfill

\end{document}